\documentclass[12pt]{article}
\usepackage{amsmath, amssymb}
\input epsf.tex

\renewcommand{\baselinestretch}{1}
\usepackage{amssymb, amsmath, amsopn, amsthm}

\pdfoutput=1

\usepackage{bm,amsfonts,array,calc}
\usepackage{epsfig}
\usepackage{graphicx}
\usepackage{color}
\usepackage[colorlinks=false]{hyperref}

\global\def\draftcontrol{0}

   \def\versionno{Higher Derivative Brane Couplings from String Amplitudes}

\catcode`\@=11

\expandafter\ifx\csname
draftcontrol\endcsname\relax\global\def\draftcontrol{0} \fi

{\count255=\time\divide\count255 by 60
\xdef\hourmin{\number\count255} \multiply\count255
by-60\advance\count255 by\time
\xdef\hourmin{\hourmin:\ifnum\count255<10 0\fi\the\count255}}
\def\draftdate{\number\month/\number\day/\number\year\ \ \ \hourmin }

\newcommand\makepapertitle{\par

  \begingroup
    \renewcommand\thefootnote{\@fnsymbol\c@footnote}%
    \def\@makefnmark{\rlap{\@textsuperscript{\normalfont\@thefnmark}}}%
    \long\def\@makefntext##1{\parindent 1em\noindent
            \hb@xt@1.8em{%
                \hss\@textsuperscript{\normalfont\@thefnmark}}##1}%
     \newpage
     \global\@topnum\z@   
     \@makepapertitle
     \thispagestyle{empty}\@thanks
  \endgroup
  \setcounter{footnote}{0}%
  \global\let\thanks\relax
  \global\let\makepapertitle\relax
  \global\let\@makepapertitle\relax
  \global\let\@thanks\@empty
  \global\let\@author\@empty
  \global\let\@date\@empty
  \global\let\@title\@empty
  \global\let\title\relax
  \global\let\author\relax
  \global\let\date\relax
  \global\let\and\relax
  \def\version{\let\version\@version\@gobble}
}
\def\@makepapertitle{%
  \newpage
   \ifnum\draftcontrol=1{}
   \version\versionno
   \vskip 5em%
   \else
   \hfill\hbox to 3cm {\parbox{4cm}{\@pubnum}\hss}%
   \vskip 5em%
   \fi
   \begin{center}%
   \let \footnote \thanks
      {\hskip -0\textwidth \hbox to 1\textwidth%
        {\centerline{\Large\bf{\noindent\@title}}}}%
     \vskip 2em%
     {\normalsize
       \lineskip .5em%
       \begin{tabular}[t]{c}%
         \@author
       \end{tabular}\par}%
     \vskip 1em%
     {\@bstract}%
     \end{center}%
     \vfill
     \@date%
     \vskip 1.5em%
     \noindent
     \rule{12em}{.02em}\par\noindent
     \@email%
   \par
}

\gdef\@pubnum{}
\def\pubnum#1{%
  \gdef\@pubnum{#1}}

\gdef\@bstract{}
\def\Abstract#1{%
  \gdef\@bstract{%
   \parbox{\textwidth-0pc}{%
   \centerline{\bf Abstract}\penalty1000
   \noindent
   \renewcommand\baselinestretch{1.0}
   {#1}}}
}

\gdef\@email{}
\def\email#1{%
   \gdef\@email{%
  {\small Email: {\tt #1}}}
}

\def\ps@paper{\let\@mkboth\@gobbletwo%
     \ifnum\draftcontrol=1
        \def\@oddfoot{\hbox to \textwidth{\tiny \versionno \hfil\tiny\draftdate}%
        \hskip -\textwidth \hbox to \textwidth{\hfil\rm\thepage\hfil}}%
     \else\def\@oddfoot{\hbox to \textwidth{\hfil\rm\thepage\hfil}}
     \fi
     \let\@evenfoot\@oddfoot
}

\newenvironment{acknowledgments}{%
\vskip 3.25ex
\noindent {\bf Acknowledgments}
}

\def\@version#1{\ifnum\draftcontrol=1
\typeout{}\typeout{#1}\typeout{}
\vskip3mm\centerline{\hbox{\fbox{\normalsize{\tt DRAFT -- #1 -- }
                   {\draftdate}}}}\vskip3mm
\fi}
\let\version\@version
\long\def\eqlabel#1{\ifnum\draftcontrol=1
                    \tag@false  
                    \tag*{(\theequation) \hbox to -0.2cm{\hspace{0cm}\small{#1}\hss}}
                    \refstepcounter{equation}
                    \edef\@currentlabel{\theequation}
                    \ltx@label{#1}          
                    \else
                    \label{#1}
                    \fi
                    }
\let\st@bibitem\@bibitem
\let\st@lbibitem\@lbibitem
\ifnum\draftcontrol=1
  \def\@bibitem#1{%
    \st@bibitem{#1}\a@@label{#1}\ignorespaces}
  \def\@lbibitem[#1]#2{%
    \st@lbibitem[#1]{#2}\a@@label{#2}\ignorespaces}
  \def\a@@label#1{%
    \gdef\a@lab{\smash{\normalfont\small#1}}
    \ifvmode
      \if@inlabel
        \global\setbox\@labels\hbox{%
          \llap{\a@lab\let\a@lab\relax
                \kern\@totalleftmargin\kern\marginparsep}%
          \box\@labels}%
      \fi
    \fi}
\fi

\renewcommand\baselinestretch{1.25}
\renewcommand\section{\@startsection {section}{1}{\z@}%
                                   {-3.5ex \@plus -1ex \@minus -.2ex}%
                                   {2.3ex \@plus.2ex}%
                                   {\normalfont\large\bfseries}}
\renewcommand\subsection{\@startsection{subsection}{2}{\z@}%
                                   {-3.25ex\@plus -1ex \@minus -.2ex}%
                                   {1.5ex \@plus .2ex}%
                                   {\normalfont\normalsize\bfseries}}
\renewcommand\subsubsection{\@startsection{subsubsection}{3}{\z@}%
                                   {-3.25ex\@plus -1ex \@minus -.2ex}%
                                   {1.5ex \@plus .2ex}%
                                   {\normalfont\normalsize\it}}
\renewcommand\paragraph{\@startsection{paragraph}{4}{\z@}%
                                   {-3.25ex\@plus -1ex \@minus -.2ex}%
                                   {1.5ex \@plus .2ex}%
                                   {\normalfont\normalsize\bf}}
\renewcommand\subparagraph{\@startsection{subparagraph}{5}{\z@}%
                                   {-1.25ex\@plus -1ex \@minus -.2ex}%
                                   {0ex \@plus .2ex}%
                                   {\normalfont\normalsize\it}}

\numberwithin{equation}{section}

\long\def\@makecaption#1#2{%
  \vskip\abovecaptionskip
  \sbox\@tempboxa{{\bf #1:} #2}%
  \ifdim \wd\@tempboxa >\hsize
    {\small\bf #1:} {\small #2}\par
  \else
    \global \@minipagefalse
    \hb@xt@\hsize{\hfil\box\@tempboxa\hfil}%
  \fi
  \vskip\belowcaptionskip}

\renewcommand*\l@section[2]{%
  \ifnum \c@tocdepth >\z@
    \addpenalty\@secpenalty
    \addvspace{.5em \@plus\p@}%
    \setlength\@tempdima{1.5em}%
    \begingroup
      \parindent \z@ \rightskip \@pnumwidth
      \parfillskip -\@pnumwidth
      \leavevmode \bfseries
      \advance\leftskip\@tempdima
      \hskip -\leftskip
      #1\nobreak\hfil \nobreak\hb@xt@\@pnumwidth{\hss #2}\par
    \endgroup
  \fi}
\renewcommand*\l@subsection{\addvspace{.0em \@plus\p@}\@dottedtocline{2}{1.5em}{2.3em}}
\renewcommand*\l@subsubsection{\addvspace{-.2em \@plus\p@}\@dottedtocline{3}{3.8em}{3.2em}}

\definecolor{refcol}{rgb}{0.2,0.2,0.8}
\definecolor{eqcol}{rgb}{.6,0,0}
\definecolor{purple}{cmyk}{0,1,0,0}

\gdef\@citecolor{refcol} \gdef\@linkcolor{eqcol}
\def\colorlinkspurple{\gdef\@urlcolor{purple}}
\def\colorlinksblue{\gdef\@urlcolor{blue}}
\def\colorlinksred{\gdef\@urlcolor{red}}

\newcommand{\beq}{\begin{equation}}
\newcommand{\eeq}{\end{equation}}
\newcommand{\beqa}{\begin{eqnarray}}
\newcommand{\eeqa}{\end{eqnarray}}
\newcommand{\beqar}{\begin{eqnarray*}}
\newcommand{\eeqar}{\end{eqnarray*}}
\newcommand{\non}{\nonumber \\}
\newcommand{\bea}{\begin{eqnarray}}
\newcommand{\eea}{\end{eqnarray}}
\newcommand{\al}{\alpha}
\newcommand{\be}{\beta}
\newcommand{\del}{\delta}

\newcommand{\eps}{\epsilon}

\newcommand{\p}{\phi}
\newcommand{\sig}{\sigma}

\newcommand{\eg}{{\it e.g.,}\ }
\newcommand{\ie}{{\it i.e.,}\ }
\newcommand{\labell}[1]{\label{#1}} 
\newcommand{\reef}[1]{(\ref{#1})}

\newcommand\cT{{\cal T}}
\newcommand\cA{{\cal A}}

\newcommand\cL{{\cal L}}

\newcommand\Tr{{\rm Tr}}

\begin{document}

\vspace*{1cm}

\begin{center}
{\bf \Large  On Holographic Weyl Anomaly}

\vspace*{1cm}

{Mozhgan Mir\footnote{mo\_mi757@stu-mail.um.ac.ir}}\\
\vspace*{1cm}
\it Perimeter Institute for Theoretical Physics, Waterloo,\\ \it Ontario N2L 2Y5, Canada\\
\it Department of Physics, Ferdowsi University of Mashhad,\\ \it P.O. Box 1436, Mashhad, Iran
\\
\vspace{2cm}

\end{center}

\begin{abstract}
\baselineskip=18pt
We use the relation between certain diffeomorphisms in the bulk and Weyl transformations on the boundary to build the conformal structure of the metric in the presence of matter in the bulk. We explicitly obtain the conformal anomaly in any spacetime dimension d in a holographic frame work in the case of gravity coupled to the scalar fields. This way we also provide a holographic construction of the bulk spacetime metric and of the matter fields in the bulk of this spacetime out of sources that are related to conformal field theory data. With the latter, we produce an asymptotic expansion of the bulk fields and scalar fields near the boundary in terms of spacetime dimension in the context of the AdS/CFT correspondence. We work out both the gravitational and matter conformal anomalies of the boundary theory as a coefficient of the infrared logarithmic divergence that appears in the on-shell action.
\end{abstract}

\vfill
\setcounter{page}{0}
\setcounter{footnote}{0}
\newpage

\tableofcontents

\section{Introduction} \label{intro}
Holography prescribes a correspondence between a (d+1)-dimensional gravitational theory and a d-dimensional field theory referred to bulk/boundary duality \cite{tHooft:1993dr,Susskind:1995wh}. The AdS/CFT correspondence provides a precise dictionary between the bulk and boundary physics \cite{Witten:1998ads,Susskind:98} with a rather exact computational frame work \cite{Witten:1998ads,Gubser:98}. It is proposed in \cite{Witten:1998ads,Gubser:98} that the partition function of string theory on AdS spaces with determined boundary conditions for the bulk fields corresponds to a generating functional of correlation functions in the conformal field theory (CFT). String theory at low energies is understood as a supergravity theory that corresponds to the large N and strong coupling regime of the CFT \cite{Maldacena:1998ln}. The boundary value of fields is playing the role of sources for operators of the dual CFT\footnote{We only assume that the dual CFT exists and do not have any further assumption since our results only depend on the spacetime dimension of the boundary theory.}. For instance, the boundary metric couples to the boundary stress-energy tensor. Also there is a correspondence between the infrared divergences on the supergravity side and the ultraviolet divergences on the CFT side.

One may construct an asymptotically AdS space through solving the gravitational equations coupled to scalars and the scalar field equation on this manifold, by using the conformal structure at infinity and the boundary value of the scalars as initial data. However, a boundary conformal structure does not determine the bulk metric completely and one needs more CFT data to specify the solution to the higher order terms in the asymptotic expansion of the solution for instance the expectation value of the corresponding operators \cite{Papadimitriou:2004ads}. Although under certain assumption for example conformally flat bulk metrics  does yield the bulk metric up to diffeomorphisms \cite{SkenderisQ:99}. Let us emphasize that we do not have any assumption about the regularity of the solution, since we only obtain the near boundary solution, therefore we simply do not confront possible singularities in the interior of the bulk.

Any asymptotically AdS metric near the boundary in the Graham-Fefferman coordinate systems \cite{Fefferman:1995ci} can be rewritten as \cite{Graham:91,Graham :99,Grahamvol:99}
\beqa
ds^2=G_{\mu\nu}dx^{\mu}dx^{\nu}=L^2\Big(\frac{d\rho^2}{4\rho^2}+\frac{1}{\rho}g_{i j}(x,\rho)dx^{i}dx^{j}\Big),\labell{coord}
\eeqa
where\footnote{The index in the brackets indicates the half of the number of derivatives involved in that term}
\beqa
g(x,\rho)=g^{(0)}+\cdots+\rho^{\frac{d}{2}}g^{(\frac{d}{2})}+\rho^{\frac{d}{2}}h^{(\frac{d}{2})}\log\rho+\cdots.\labell{ggex}
\eeqa
The logarithmic term exists only in even dimensions. $L$ is related to the cosmological constant as $\Lambda=-\ \frac{d(d-1)}{2L^2}$. For simplicity throughout this paper we set $L=1$. Metric $G_{\mu \nu}$ in \reef{coord} has singularity at infinity. However performing a (singular) conformal transformation induces a well-defined boundary metric $g^{(0)}$ at infinity.

The $AdS_{d+1}$ space has the curvature of $G$ as
\beqa
R_{\mu \nu \rho \sigma}=(G_{\mu \rho}G_{\nu \sigma}-G_{\mu \sigma}G_{\nu \rho})+O(\rho).
\eeqa
The action of Weyl transformations of the boundary metric is realized as a subgroup of the bulk diffeomorphisms for the bulk metric. These conformal transformations on the boundary is called PBH transformations\cite{Imbimbo:2000diff}. The PBH transformation does not depend on the details of the solution of interest and therefore it gives information for any such solutions, \eg even including stringy corrections.

Infrared divergences in the gravitational sector can be seen as the cause of the conformal anomaly \cite{Henningson:1998wel}. Invariance of AdS under rescaling the radial coordinate that corresponds to dilatation in the CFT, is broken after regulating these infrared divergences. It means that a conformal transformation in the boundary theory is related to a special diffeomorphism in the bulk that preserves the form of the coordinate system \reef{coord}\cite{Imbimbo:2000diff}.

In the asymptotic solution \reef{ggex} $g^{(n)}$ with $n<\frac{d}{2}$ are determined uniquely and locally by $g^{(0)}$. $g^{(\frac{d}{2})}$ is a nonlocal quantity which is not obtained only from the boundary condition for the metric. Rather only its trace and covariant divergence are determined. However, it is determined by the vacuum expectation value of the dual stress-energy tensor. In the case of pure gravity, when the boundary dimension d is odd the coefficient $g^{(\frac{d}{2})}$ in \reef{ggex} is conserved and traceless and there is no quantum correction for the stress-energy tensor. Furthermore it is well-known that $h^{(\frac{d}{2})}$ is related to the conformal anomaly through the metric variation of the conformal anomaly \cite{Haro:2001hr}. It was shown in \cite{SkenderisQ:99} that, given a conformally flat boundary metric in which the bulk Weyl tensor vanishes, the bulk metric can be determined to all orders. Also in this case $h_{(\frac{d}{2})}$ in \reef{ggex} vanishes because it is the metric variation of a topological invariant. In addition for the standard Euclidean AdS, $g_{(n)}=0$ for $0<n<\frac{d}{2}$ \cite{Haro:2001hr}.

In the case of pure gravity, the holographic conformal anomalies for the spacetime dimensions up to six were obtained \cite{Haro:2001hr} and the holographic results agree with the known field theory calculations\cite{Henningson:1998wel}. It is well-known that in odd (boundary) dimensions, there are no conformal anomalies. We would like to build holographically the bulk metric coupled to the matter out of the CFT data and extract the associated conformal anomaly as a quantum feature of CFT by applying the holographic prescription \cite{Haro:2001hr}. The CFT data are the sources for the operators that are turned on and the corresponding expectation values.

The same discussion is applicable when one adds matter to the system. By considering scalars that correspond to marginal and relevant operators one can compute perturbatively the back reaction of the scalars to the gravitational background. The solution of the massive scalar field has the mass
\beqa
m^2=\Delta(\Delta-d),\labell{m2}
\eeqa
where $\Delta$ is the conformal dimension of the dual operator\cite{Aharony:99}. The asymptotic expansion for the scalar field near the boundary in the coordinate system \reef{coord} is in the form \cite{Haro:2001hr}
\beqa
\Phi(x,\rho)&=&\rho^{\frac{(d-\Delta)}{2}}\phi(x,\rho),\qquad\phi(x,\rho)=\phi^{(0)}+\phi^{(1)}\rho+\ldots .\labell{fiex0}
\eeqa
Because we want to use the probe approximation, we consider a relevant operator in the boundary theory. It requires in  \reef{m2}, $m^2<0$ for the bulk scalar and this approach is valid in the regime $\Delta<d$ \ \footnote{For a marginal operator with $\Delta=d$ (or $\al=0$) we will see in \reef{resclag} that the matter part of the anomaly vanishes and we left with the gravitational part.}. The leading order in the small-$\rho$ expansion of the scalar field is $\rho^{\frac{\alpha}{2}}$ with $\alpha=d-\Delta$ which appears in the mass formula \reef{m2} and
\beqa
0<\alpha<\frac{d}{2}.
\eeqa
Although in the case pure gravity for odd boundary dimensions there are no gravitational conformal anomalies, adding relevant deformations there is a possibility of having conformal anomalies in both odd and even dimensions, providing
$\Delta-\frac{d}{2}$ to be an integer, a logarithmic term appears. The logarithmic term $\psi^{(\Delta-\frac{d}{2})}$ appearing in this situation in \reef{fiex0}, is related to the matter conformal anomalies \cite{Haro:2001hr}. $\phi^{(n)}$ with $n<\Delta-\frac{d}{2}$ and $\psi_{(\Delta-\frac{d}{2})}$ are determined uniquely in terms of $g^{(0)}$ and $\phi^{(0)}$ from the scalar field equation and gravitational equations coupled to scalars that are presented in the next section.
The coefficient $\phi^{(\Delta-\frac{d}{2})}$ is not obtained by solving the field equations. However it is the leading finite part  determined by the vacuum expectation value\footnote{Global condition on the bulk metric corresponds to the choice of a particular vacuum in the conformal field theory.} of the dual operator \cite{{Klebanov:99}}
\beqa
\langle O(x)\rangle &=&(\Delta-\frac{d}{2})\phi_{(\Delta-\frac{d}{2})}(x).\nonumber
\eeqa
In this paper we follow two parallel processes to construct the holographic conformal anomaly in the presence of  matter. In section \ref{asym} we discuss how to find the anomaly by solving equations of motion. In section \ref{PBH} we use the PBH transformations that give us the covariant form of the anomaly in the cases we are interested but the overall factors are still undetermined. In section \ref{EOM} first we obtain an asymptotic solution in the case of gravity coupled to a free scalar by choosing specific dimensions $d=4,6$ for a given boundary metric and Dirichlet value of the scalar field. We use these solutions to figure out the infrared divergences of both gravitational and matter sectors.
Then in subsections \ref{EOMdk1} and \ref{EOMdk2} we generalize this approach to a general dimension and compute the conformal anomaly, in terms of d with given boundary conditions, as local polynomial functions of two and four derivative terms built of the Ricci tensor and the Ricci scalar of the metric ${g_{(0)}}_{i j}$ and the scalar field in the boundary value $\phi^{(0)}$ and their covariant derivatives. In the last three subsections, we exhibit results for the insertion of four scalar fields.
\section{Asymptotic AdS gravity coupled to scalar fields} \label{asym}
The Einstein-Hilbert action for a gravity theory with $d+1$ dimensions on a manifold $M$ and the d-dimensional boundary\footnote{In the following, d+1-dimensional indices are shown with Greek indices $\mu,\nu,\cdots$, d-dimensional ones with Latin ones $i,j,\cdots$. $R_{i j}[G]$ indicates the Ricci tensor of the metric G, etc.} $\partial M$ is
\beqa
S_{gr}&\!\!\!\!=\!\!\!\!&\frac{1}{16\pi G^{d+1}_N}[\int_M d^{d+1}x \sqrt{G}(R[G]+2\Lambda)-\int_{\partial M} d^{d}x \sqrt{\gamma}2K],\labell{graction}
\eeqa
where $K$ is the trace of the second fundamental form and $\gamma$ is the induced metric on the boundary. The boundary term makes the variational problem with Dirichlet boundary conditions be well-defined \cite{Gibbons:77}. Here we work with Euclidean signature of the metric.

The on-shell gravitational action diverges because of the spacetime volume is infinite and also the singularity in the metric at the boundary as was mentioned previously. To regulate the theory, we cut off the spacetime at $\rho=\epsilon$ for some small parameter $\epsilon$ near the boundary. In the action \reef{graction} the bulk integral is calculated in the region $\rho\geq\epsilon$ and the boundary term at $\rho=\epsilon$. The divergences in the action are in the form of the $1/\epsilon^k$ poles and in the case of even (boundary) dimensions there is also a logarithmic divergence \cite{Henningson:1998wel}
\beqa
S_{gr,reg}&\!\!\!\!=\!\!\!\!&\frac{1}{16\pi G^{d+1}_N}\int d^{d}x \sqrt{ g_{(0)}}\Big(\epsilon^{-\frac{d}{2}}a_{(0)}+\epsilon^{-\frac{d}{2}+1}a_{(1)}+\cdots+\epsilon^{-1}a_{(\frac{d}{2}-1)}-\log \epsilon \ a_{(\frac{d}{2})} \Big)\nonumber\\
&&\left.\qquad\qquad\qquad\qquad\qquad\qquad\qquad\qquad\qquad\qquad\qquad\qquad\qquad+O(\epsilon^0),\right.\labell{regraction}
\eeqa
where the coefficients $a_{(n)}$\footnote{ We will see later in the presence of scalar fields, the coefficients $a_{(n)}$ also depend on the scalar field and its covariant derivatives as well.
} are given by local covariant combinations of the $g_{(0)}$ and its curvature tensor.

The renormalized action is obtained by subtracting the divergent terms and then removing the regulator, \ie $\epsilon\rightarrow 0 $. The terms proportional to the negative powers of $\epsilon$ in the regularized action are invariant under the scale transformations
\beqa
\delta g^{(0)}=2\delta \sigma g^{(0)},\qquad \delta \epsilon=2\delta \sigma \epsilon .\nonumber
\eeqa
The $\log \epsilon$ term transforms as $\log \epsilon\rightarrow \log \epsilon+\log(2\delta\sigma)$. The variation from the latter does not depend on $\epsilon$ and should be canceled by the variation of the finite terms such that the total action \reef{regraction} remains invariant under the transformation. Hence, the $\log \epsilon$ term gives the conformal anomaly $\cA$ of the renormalized theory on the boundary
\beqa
S_{\lg,g}=-\ \frac{16\pi G^{d+1}_N}{2}\int d^d x \sqrt{g_{(0)}} \cA .\labell{anomalydef}
\eeqa
Generally the boundary conditions for the metric and scalar fields are arbitrary fields because they are regarded as sources for the boundary operators and we take functional derivatives with respect to them in the corresponding conformal field theory. Also solving the equations of motion and applying the boundary conditions will lead us to the information in the boundary theory such as anomaly. The gravitational field equation in the presence of matter reads
\beqa
R_{\mu \nu}-\frac{1}{2}(R+2\Lambda) G_{\mu \nu}=-8\pi G_N T_{\mu \nu},\labell{Einsceq}
\eeqa
with appropriate Dirichlet boundary conditions.

In order to solve \reef{Einsceq} with a given  conformal structure at infinity, we work in the coordinate system \reef{coord} introduced by Fefferman and Graham \cite{Fefferman:1995ci}. In the case of relevant deformation on the boundary theory, the metric expansion becomes \cite{Hung:2011scc}
\beqa
g_{i j}(x,\rho)&=&\sum_{n=0}^{N-1}\bigg(\rho^n g^{(n)}_{i j}(x)+\sum_{m=2}^{\infty}\rho^{n+m \frac{\al}{2}} g^{(n+m \frac{\al}{2})}_{i j}(x)\bigg)\nonumber\\
&&\left.+\rho^{\frac{d}{2}}\sum_{n=0}^{N-1}\bigg(\rho^n g^{(\frac{d}{2}+n)}_{i j}(x)+\sum_{m=2}^{\infty}\rho^{n+m \frac{\al}{2}} g^{(\frac{d}{2}+n+m \frac{\al}{2})}_{i j}(x)\bigg),\right.\labell{g ex}
\eeqa
where we assume $\frac{\al}{2}=\frac{N}{M}$ which M, N are relatively prime.

For even dimensions, a logarithmic term appearers in the expansion as follows \cite{Hung:2011scc}:
\beqa
g_{i j}(x,\rho)&=&\sum_{n=0}^{N-1}\bigg(\rho^n g^{(n)}_{i j}(x)+\sum_{m=2}^{\infty}\rho^{n+m \frac{\al}{2}} g^{(n+m \frac{\al}{2})}_{i j}(x)\bigg)\nonumber\\
&&\left.+\rho^{\frac{d}{2}}\log \rho \sum_{n=0}^{N-1}\bigg(\rho^n f^{(\frac{d}{2}+n)}_{i j}(x)+\sum_{m=2}^{\infty}\rho^{n+m \frac{\al}{2}} f^{(\frac{d}{2}+n+m \frac{\al}{2})}_{i j}(x)\bigg).\right.\labell{gex2}
\eeqa
We wish to study how the bulk spacetime with a relevant deformation in the boundary theory is constructed holographically . Here we discuss a bulk scalar field but a similar approach is applicable to other kinds of matter. The action for a free massive scalar is given by
\beqa
S_{M}&\!\!\!\!=\!\!\!\!&\frac{1}{2}\int d^{d+1}x \sqrt{G}(G^{\mu \nu}\partial_{\mu}\Phi\partial_{\nu}\Phi+m^2 \Phi^2),\labell{scaction}
\eeqa
where the metric $G_{\mu \nu}$ has an expansion of the form \reef{gex2}.

Similar to the case of pure gravity, the theory is regularized by integrating the bulk $\rho\geq\epsilon$ in the action \reef{scaction}\cite{Viswanathan:98,Freedman:98}
\beqa
S_{M,reg}&\!\!\!\!=\!\!\!\!&\int d^{d}x \sqrt{det \ g_{(0)}}\Big(\epsilon^{\alpha-\frac{d}{2}}a^M_{(0)}+\epsilon^{\alpha-\frac{d}{2}+1}a^M_{(1)}+\cdots+\epsilon \ a^M_{(\alpha-\frac{d}{2}+1)}-\log \epsilon \ a^M_{(\alpha-\frac{d}{2})} \Big)\nonumber\\
&&\left.\qquad\qquad\qquad\qquad\qquad\qquad\qquad\qquad\qquad\qquad\qquad\qquad\qquad+O(\epsilon^0),\right.\labell{rescaction}
\eeqa
where the matter conformal anomaly is given by the coefficient of the logarithmic term and is added to the formula \reef{anomalydef} to obtain the whole contribution. The form of the expansion of the scalar field $\Phi$ is \cite{Hung:2011scc}
\beqa
\Phi(x,\rho)=\rho^{\frac{\alpha}{2}}\phi(x,\rho)&\!\!\!\!=\!\!\!\!&\rho^{\alpha/2}\sum_{n=0}^{N-1}\sum_{m=0}^{\infty}\rho^{n+m \frac{\al}{2}} \phi^{(n+m \frac{\al}{2})}(x)\nonumber\\
&&\left.+\rho^{\frac{(d-\alpha)}{2}} \sum_{n=0}^{N-1}\sum_{m=0}^{\infty}\rho^{n+m \frac{\al}{2}} \phi^{(\frac{d}{2}-\alpha+n+m \frac{\al}{2})}(x).\right.\labell{sumscex}
\eeqa
By applying this constraint, we take operator with a particular conformal dimension $\Delta$ to get logarithmic term, \ie
\beqa
\Delta=\frac{d}{2}+k,\qquad k=0,1,\cdots.\nonumber
\eeqa
The scalar field equation $(-\Box_G+m^2)\Phi=0$ by using \reef{fiex0} becomes
\beqa
[-\alpha\partial_{\rho}\log g \phi +2(d-2\alpha-2)\partial_{\rho}\phi-\frac{1}{\sqrt{g}}\partial_i(\sqrt{g}{g^{-1}}^{i j}\partial_j\phi)]+\rho[-2\partial_{\rho}\log g \partial_{\rho}\phi-4\partial^2_{\rho}\phi]=0.\nonumber\\ \labell{seq}
\eeqa
where we used the relation $m^2=\alpha(\alpha-d)$ for the mass and $\Box_G \Phi=\frac{1}{\sqrt{G}}\partial_\mu(\sqrt{G}G^{\mu \nu}\partial_{\nu}\Phi)$.

Through solving the equations of motion, one can determine recursively $\phi_{(n)}$, $n<\frac{d}{2}-k$. For given $\phi_{(0)}$, when $\frac{d}{2}-k$ is integer, at this order the coefficient of $\phi^{(d/2-k)}$ becomes zero and it is necessary to introduce a logarithmic term at this order, now the expansion of $\Phi$ reads
\beqa
\Phi(x,\rho)=\rho^{\frac{\alpha}{2}}\phi(x,\rho)&\!\!\!\!=\!\!\!\!&\rho^{\alpha/2}\sum_{n=0}^{N-1}\sum_{m=0}^{\infty}\rho^{n+m \frac{\al}{2}} \phi^{(n+m \frac{\al}{2})}(x)\nonumber\\
&&\left.+\rho^{\frac{(d-\alpha)}{2}} \log\rho\sum_{n=0}^{N-1}\sum_{m=0}^{\infty}\rho^{n+m \frac{\al}{2}} \psi^{(\frac{d}{2}-\alpha+n+m \frac{\al}{2})}(x).\right.\labell{sumscex}
\eeqa
The coefficient of the logarithmic term $\psi^{(\Delta-\frac{d}{2})}$ is obtained but $\phi^{(\Delta-\frac{d}{2})}$ remains undetermined. However, it is known \cite{Balasubramanian:98,Klebanov:99} that the latter is specified by the expectation value of the dual operator. Since the action does not include interactions for the scalar, we can expect $\phi^{(\Delta-\frac{d}{2})}$ to be linear in $\phi^{(0)}$.

Now we proceed to obtain solutions to Einstein's equations coupled to scalar fields. The total action is given by the summation of \reef{graction} and \reef{scaction}
\beqa
S&=&S_{gr}+S_{M}.\labell{toaction}
\eeqa
The gravitational equations of motion\footnote{Notice that if $\al<0$ (or $\Delta>d$) the right hand side of equations diverges near the boundary, because for $\al<0$ operators are irrelevant. We only consider  $\al\geq 0$ (or $\Delta\leq d$) that gives marginal or relevant operators and equations can be calculated perturbatively.} \reef{Einsceq} in the coordinate system \reef{coord} by making use of the expansion of the scalar field \reef{fiex0} read \cite{Haro:2001hr}
\beqa
\rho\left(2 g_{i j}^{\prime \prime}-2(g^{\prime} g^{-1}g^{\prime})_{i j}\right.&+&\left.Tr(g^{-1}g^{\prime}){g_{i j}^{\prime}}\right)-R_{ij}(g)-(d-2){g_{i j}^{\prime}}
-\Tr(g^{-1}g^{\prime}){g}_{i j}=\nonumber\\
&=&-8 \pi G^{d+1}_N\rho^{\alpha-1}\left(\frac{\alpha(\alpha-d)}{d-1}\phi^2g_{i j}+\rho\ \partial_i\phi\partial_j\phi\right)\labell{greom} \\
\nabla_i \Tr(g^{-1}g^{\prime})-\nabla^j g^{\prime}_{i j}&=&-16 \pi G^{d+1}_N \rho^{\alpha-1}
\left(\frac{\alpha}{2}\phi\ \partial_i\phi+\rho\ \partial_{\rho}\phi \partial_{i}\phi\right)
 \non
\Tr(g^{-1} {g^{\prime \prime}})-\frac{1}{2}\Tr(g^{-1}g^{\prime}g^{-1}g^{\prime})&=&-16\pi G^{d+1}_N \rho^{\alpha-2}\left(\frac{\alpha d(\alpha-1)}{4(d-1)}\phi^2 +\alpha \rho\ \phi \partial_{\rho}\phi+\rho^2 (\partial_{\rho}\phi)^2
 \right)\nonumber
 \eeqa
where prime presents differentiation with respect to $\rho$ (with $\rho$ considered an extra parameter, rather than a coordinate), $\nabla_i$ is the covariant derivative built from the metric g, and $ R_{i j}(g)$ is the Ricci tensor of g.

These equations are solved order by order in $\rho$.
First from the equation in \reef{greom} the coefficients $g_{(n)}$ for $n\neq \frac{d}{2}$ in terms of $g_{(0)}$ are obtained. The expression for $g_{(n)}$ becomes singular when $n=\frac{d}{2}$. The trace and divergence of $g_{(n)}$ for any n are determined from the last two equations in \reef{greom}. Since the trace of $g_{(\frac{d}{2})}$ is related to the vacuum expectation value of the trace stress tensor, if the former is zero we conclude that the stress energy tensor does not receive quantum effects, otherwise it detects the appearance of the conformal anomaly that is obtainable from both coefficients of the logarithmic terms, $h_{(\frac{d}{2})}$ and $\psi_{(\frac{d}{2}-\alpha)}$.

From the regulated on-shell action,the value of the bulk integral in \reef{toaction} gives a logarithmic divergence that is where the anomaly emerges \cite{Haro:2001hr}
\beqa
S_{reg}(bulk)&=&\int_{\rho\geq\epsilon} d\rho\ d^{d}x \sqrt{G}[ \frac{d}{8\pi G_N^{(d+1)}}-\frac{m^2}{d-1}\Phi^2]\labell{rescaction}\\
&=&\int_{\rho\geq\epsilon} d\rho\ d^{d}x \frac{1}{\rho}\sqrt{g(x,\rho)}[ \frac{d}{16\pi G_N^{(d+1)}}\rho^{-d/2}-\frac{m^2}{2(d-1)}\phi^2(x,\rho)\rho^{-k}].\nonumber
\eeqa
where $k=\frac{d}{2}-\alpha$, from above formula it is concluded even if d is odd there is a possibility of having conformal anomaly as k is a positive integer.
\section{Anomaly by the PBH transformation approach} \label{PBH}
AdS/CFT duality states that the conformal transformations in the boundary are emerged from the specific diffeomorphism that preserves the form of the metric \reef{coord} \cite{Gubser:99}. Consider a coordinate transformation that keeps the form of the metric in FG gauge \reef{coord}
\beqa
\rho=\rho^{\prime}(1-2\sigma(x^{\prime}))\qquad x^i=x^{\prime i}+a^i(x^{\prime},\rho^{\prime}).\nonumber
\eeqa
The $a^i(x^{\prime},\rho^{\prime})$ are defined providing of the form invariance of the metric. To leading order in $\sigma$, with boundary condition $a^i(x,\rho=0)$, it is given by
\beqa
a^i(x,\rho)=\frac{L^2}{2}\int_0^{\rho}d\rho^{\prime}g^{i j}(x,\rho^{\prime})\partial_j\sigma(x).\labell{axro}
\eeqa
The infinitesimal diffeomorphism transformation of $g_{i j}(x,\rho)$ under the PBH transformation \cite{Imbimbo:2000diff} is
\beqa
\del g_{i j}(x,\rho)=2 \sig (1-\rho \partial_{\rho})g_{i j}(x,\rho)+\nabla_i a_j(x,\rho)+\nabla_j a_i(x,\rho),\labell{varyg}
\eeqa
where the covariant derivatives are constructed from the metric $g_{i j}(x,\rho)$. The first term is a homogenous term and its coefficient determines the conformal dimension. To see this, regard the power series expansion of the metric in the gravity theory coupled to scalars in the vicinity of $\rho=0$
\beqa
g(x,\rho)&\!\!\!\!=\!\!\!\!
&g^{(0)}+g^{(1)}\rho+\ldots+g^{(n)}\rho^n
 +\rho^{\al}\Big(g^{(\al)}
+g^{(\al+1)}\rho+g^{(\al+2)}\rho^2+\ldots\Big), \labell{gex00}
\eeqa
then one gets a homogenous scaling term for each coefficient as
\beqa
\delta g^{(n)}=-2 \sigma(x)(n-1)g^{(n)}+\cdots ,\nonumber
\eeqa
it implies that ${g_{(n)}}_{i j}$ has the conformal dimension $2(n-1)$. In particular, ${g_{(0)}}_{i j}$ has the conformal dimension $-2$. To find how different components of the metric  transform, we expand $a^i(x,\rho)$ in powers of $\rho$
\beqa
a=a^{(1)}\rho+a^{(2)}\rho^2+\ldots+a^{(n+1)}\rho^{n+1}+\rho^{\al}
\Big(a^{(\al+1)}\rho+a^{(\al+2)}\rho^{2}+\ldots\Big),\labell{aex0}
\eeqa
where pertaining to the boundary condition we set $a_{(0)}^i(x)$ to be zero. We substitute above expansions into eq. \reef{varyg} and find\footnote{The symmetrization bracket is defined as $A_{(i} B_{j)}=1/2(A_i B_j+B_i A_j)$.}
 \beqa
\delta {g_{i j}^{(0)}}&=&2\sigma \ {g_{i j}^{(0)}},\labell{dga12}\\ \delta {g_{i j}^{(1)}}&=&a_{(1)}^k \partial_k {g_{i j}^{(0)}}+\partial_{(i}  a_{(1)}^k {g_{j) k}^{(0)}},\nonumber\\
\delta {g_{i j}^{(\alpha)}}&=&-2\sigma (\alpha-1) {g_{i j}^{(\alpha)}},\nonumber\\
\del g^{(\al+1)}_{i j}&=& -2 \sig\al g^{(\al+1)}_{i j}+a^k_{(1)}\partial_k g^{(\al)}_{i j}+a^k_{(\al+1)}\partial_k g^{(0)}_{i j}+2\partial_{(i} a^k_{(1)} g^{(\al)}_{j) k}+2\partial_{(i} a^k_{(\al+1)} g^{(0)}_{j) k},\nonumber\\
\del g^{(\al+2)}_{i j} &=& -2 \sig (\al+1)g^{(\al+2)}_{i j} +a^k_{(\al+2)}\partial_k g^{(0)}_{i j} +a^k_{(\al+1)}\partial_k g^{(1)}_{i j} +a^k_{(1)}\partial_k g^{(\al+1)}_{i j} +a^k_{(2)}\partial_k g^{(\al)}_{i j}
 \nonumber\\
&&\left.+2\partial_{(i} a^k_{(\al+2)} g^{(0)}_{j) k} +2\partial_{(i} a^k_{(\al+1)} g^{(1)}_{j) k} +2\partial_{(i} a^k_{(1)} g^{(\al+1)}_{j) k} +2\partial_{(i} a^k_{(2)} g^{(\al)}_{j) k},  \right.\nonumber
\eeqa
where
\beqa
a_{(1)}^{i}&=& \frac{L^2}{2 }({g^{(0)}}^{-1})^{i j}\partial_j \sig, \qquad a^i_{(2)}=-\frac{L^2}{4}{g_{(1)}^{-1}}^{i j}\partial_j \sigma,\nonumber\\
a_{(\be)}^{i}&=&-\ \frac{L^2}{2 \be}{({g^{(0)}}^{-1}g^{(\be-1)}{g^{(0)}}^{-1})}^{i j}\partial_j \sig .\labell{abeta}
\eeqa
where $\be=\al+1,\al+2,\ldots$ \footnote{Note that by referring to \reef{axro}, $\be$ starts from $\al+1$.}. Also for utilization at some point subsequently we have
\beqa
  \delta R= -2 \sigma R-2(d-1)\square \sigma,\qquad
 \delta R_{i j}= -(d-2) \nabla_i\nabla_j\sigma -g_{i j}\Box \sigma ,\labell{Rtrans}
 \eeqa
where $\Box$ exhibits the Laplacian. Furthermore we acquire these relations by taking advantage of the PBH transformations by noticing the transformation of components of the scalar field under transformations. The series expansion of the scalar field is in the following form:
\beqa
\Phi(x,\rho)&=&\rho^{\frac{\alpha}{2}}\Big(\phi^{(0)}+\phi^{(1)}\rho+\phi^{(2)}\rho^2+\ldots \Big)
+\rho^{\frac{(d-\alpha)}{2}}\Big(\phi^{(\frac{d}{2}-\alpha)}+\ldots\Big),\labell{fiex00}
\eeqa
therefore, we get
 \beqa
\delta\phi^{(0)}&=&-\alpha \sigma  \phi^{(0)},\qquad
\delta\phi^{(1)}=-(\alpha+2) \sigma  \phi^{(1)}+a^k_{(1)}\partial_k\phi^{(0)}\labell{fi012}\\
\delta\phi^{(2)}&=&-(\alpha+4) \sigma  \phi^{(2)}+a^k_{(1)}\partial_k\phi^{(1)}+a^k_{(2)}\partial_k\phi^{(0)},\nonumber
\eeqa
also
\beqa
\delta\phi^{(\alpha)}=-3\alpha \sigma   \phi^{(\alpha)},\qquad
\delta\phi^{(\alpha+1)}=-(3\alpha+1) \sigma   \phi^{(\alpha+1)}+a^k_{(\alpha+1)} \partial_k\phi^{(0)}+a^k_{(1)} \partial_k\phi^{(\alpha)},\nonumber
\eeqa
where the definitions for $a^k_{(M)}$, $M=1,2,\be$ are given in \reef{abeta}. From the last two terms we see in general the conformal dimension of $\phi^{(n)}$ is $\alpha+2n$.

Now we want to take advantage of the PBH transformations and determine the explicit form of the different components of the metric and the scalar field expansions. For this purpose, we write down the most covariant combination using the curvature tensor, the scalar field and their covariant derivatives such that each term in this combination with an arbitrary coefficient has the desired conformal dimension.
Therefore, by understanding the PBH transformations of any component given in \reef{dga12} and \reef{fi012}, we make a comparison with the conformal transformation of the mentioned combination. The equivalence of inhomogeneous parts of transformations in two sides specifies the unknown coefficients. For example the covariant expressions for $g^{(1)}$ and $g^{(2)}$ under the PBH transformations become \footnote{Notice that our convention for the curvature tensor differs from one given in \cite{Haro:2001hr} in a minus sign.}
\beqa
g^{(1)}_{ij}&=&-\ \frac{L^2}{d-2}\bigg(R^{(0)}_{ij}-\frac{R\ g^{(0)}_{ij}}{2(d-1)}\bigg),\labell{g1,g2}\\
g^{(2)}_{i j}&=&c_1 L^4 C_{ k l m n}C^{ k l m n}g^{(0)}_{i j}+c_2 L^4 C_{ i k l m }{C_j}^{ k l m}\nonumber\\&&\left.+\ \frac{L^4}{d-4}\bigg(\frac{1}{8(d-1)}\nabla_i \nabla_j R-\frac{1}{4(d-2)}\square R_{i j}+\frac{1}{8(d-1)(d-2)}\square R g^{(0)}_{i j}\right.\nonumber\\&&\left.
-\ \frac{1}{2(d-2)} R^{k l}R_{i k j l}+\frac{d-4}{2(d-2)^2} {R_i}^k R_{ j k}+\frac{1}{(d-1)(d-2)^2} R R_{i j }\right.\nonumber\\&&\left.
+\ \frac{1}{4(d-2)^2} R^{k l} R_{k l}g^{(0)}_{i j}-\frac{3 d}{16 (d-1)^2 (d-2)^2} R^2 g^{(0)}_{i j}\bigg),\right.\nonumber
\eeqa
we see later there are possible contributions coming from the scalar field (see\reef{g1d4} and \reef{g2d6}).

Lets define a Lagrangian by using the regulated bulk action \reef{rescaction} that gives only the logarithmic contribution to the action with one dimension higher
\beqa
\cL=\int_{\rho\geq\epsilon} d\rho\  \frac{1}{\rho}\sqrt{g(x,\rho)}[ \frac{d}{16\pi G_N^{(d+1)}}\rho^{-\frac{d}{2}}+\frac{\alpha(d-\alpha)}{2(d-1)}\phi^2(x,\rho)\rho^{-k}],\labell{resclag}
\eeqa
where $k=\frac{d}{2}-\alpha$ and we have inserted the definition \reef{m2}.

The result of the integration has a $-\log \epsilon$ part that based on the discussion above \reef{anomalydef}, its coefficient with multiplication of a factor of $-2$ gives us the anomaly
From the regulated action for the metric $G$ and the scalar field $\Phi$. Referring to the logarithmic divergences in \reef{rescaction} and the discussion above \reef{anomalydef}, one finds
\beqa
\cA=-2(\frac{d}{16\pi G_N^{(d+1)}}a_{(\frac{d}{2})}+a^M_{(\alpha-\frac{d}{2})}).\labell{totalanomaly}
\eeqa
The gravitational part of the anomaly appears at  order of $\frac{d}{2}$ in the expansion of $\sqrt{g(x,\rho)}$ and the matter anomaly occurs at order of $k$ in the expansion of $\sqrt{g(x,\rho)} (\phi(x,\rho))^2$. In these cases the integration in \reef{resclag} gives us a logarithmic divergence. In order to expand the square root, we use the general formula for expansion of the square root of determinant of a typical matrix $A=1+B$
\beqa
\sqrt{det(1+B)}=1+\frac{1}{2}\Tr B-\frac{1}{4}\Tr B^2+\frac{1}{8}(\Tr  B)^2-\frac{1}{8}\Tr B\ \Tr B^2+\frac{1}{6}\Tr B^3+\frac{1}{48}(\Tr B)^3+\ldots \nonumber\\\labell{detex}
\eeqa
where $B$ is obtained by
\beqa
B&=&\rho g^{(-1)}_{(0)}g^{(1)}+\ldots+\rho^n  g^{(-1)}_{(0)}g^{(n)}+\rho^{\al}  g^{(-1)}_{(0)}g^{(\al)}+\rho^{\al+1}  g^{(-1)}_{(0)}g^{(\al+1)}+\ldots.\labell{B}
\eeqa
where n is an integer. The number of derivatives and scalars determine that in which order a logarithmic term appears, as we will see in the next sections. However, in above expansion we do not consider a logarithmic term since in the integrand \reef{resclag}, only polynomial terms produce the contributions to the anomaly, even if the definition of the metric and scalar components inside \reef{B} at the certain orders includes the logarithmic term.
\subsection{k=1, general d} \label{k=1pbh}
Here we want to attain the explicit forms of $g^{(\al)}$ and $g^{(\al+1)}$  by considering the most general covariant combinations with the conformal dimensions $2(\alpha-1)$ and $2\alpha$, respectively, we need to have two scalar fields in each term, as result one has
\beqa
g^{(\alpha)}_{ij}&=&C_1 {(\p^{(0)})}^2  g^{(0)}_{ij},\nonumber\\
g^{(\al+1)}_{i j}&=&C_1 L^2 \bigg(b_1 {(\p^{(0)})}^2 R_{i j}+b_2\ g^{(0)}_{ij}{(\p^{(0)})}^2 R+b_3\ \partial_i \p^{(0)}\partial_j \p^{(0)}+b_4\ \nabla_i \nabla_j{(\p^{(0)})}^2\nonumber\\
&+& b_5\ g^{(0)}_{ij}\Box{(\p^{(0)})}^2+b_6\ g^{(0)}_{ij}\nabla_k\p^{(0)}\nabla^k \p^{(0)} \bigg).\labell{ga1}
\eeqa
Notice that from the holographic principle, the Ricci tensor and the Ricci scalar are defined in terms of boundary value of the metric $g^{(0)}$. The value of $C_1$ is specified by solving the equations of motion (see \reef{C1}). For $\al=\frac{d}{2}-1$ the coefficients  $b_1,\ldots,b_4$ are characterized  in terms of $b_5$ and $b_6$
\beqa
b_1&=& -\frac{1}{2}-\frac{1}{d}+ (2 b_5+b_6)(1-\frac{d}{2}), \nonumber\\
b_2 &=& \frac{2 + (1 + b_6 (d-2)) d}{4 (d-1) d},\nonumber\\
b_3 &=& -2(1+\frac{1}{d}) + 2(1-d) (2 b_5 + b_6), \nonumber\\
b_4&=& \frac{1}{2} \bigg((d-2)(2   b_5  + b_6 )+1\bigg).\labell{b1b4}
\eeqa
That is, the expression \reef{ga1} for any value of $b_5$ and $b_6$ is conformally invariant. It turns out  for particular dimensions they may not be determined by equations of motion
The BPH transformation approach is accurate  only for polynomial expansions, but $b_5$ and $b_6$  are singular at order the anomaly appears and we need the logarithmic term in the expansion. However, there is another method to attain them at this order of equations of motion through performing the PBH transformation for $d=2n+\eps$ and finding residual values by considering $\log\rho$ in the equations of motion as the integration of $\rho^{-1-\eps}$. Nevertheless we do not follow this approach in view of the fact that we get rid of $b_5$ and $b_6$ in the normalized form of the anomaly as it is demonstrated shortly in \reef{anomalyratio}.

Because of the existence or the factor of $\rho^{-\frac{d}{2}}$ in the first term of the lagrangian \reef{resclag}, it turns out in order to find anomaly in the case of $k=1$ only  the order $\al+1$ in the expansion of the square root of the determinant of the metric $g$ contributes. In this case the explicit form for the expansion of the metric, according to \reef{gex2}, is
\beqa
g(x,\rho)&\!\!\!\!=\!\!\!\!
&g^{(0)}{(x)}+g^{(1)}{(x)}\rho+\ldots+g^{(n)}{(x)}\rho^n\nonumber\\
&&\left.+\rho^{\al}\Big(g^{(\al)}{(x)}
+g^{(\al+1)}{(x)}\rho+h^{(\al+1)}{(x)}\rho\
\log \rho+\ldots\Big),\right.\labell{gexa1}
\eeqa
where $n$ is an integer number and $\alpha=\frac{d}{2}-k$. Notice that the logarithmic term only appears when $\alpha+1=\frac{d}{2}$ is an integer. The expansion of the scalar field \reef{sumscex} that is applicable here reads
\beqa
\Phi(x,\rho)&=&\rho^{\alpha/2}\Big(\phi^{(0)}(x)+\phi^{(1)}(x)\rho+\phi^{(2)}(x)\rho^2+\ldots\Big)\nonumber\\
&&\left.+\rho^{(d-\alpha)/2}\Big(\phi^{(d/2-\alpha)}(x)+\psi^{(d/2-\alpha)}(x)\log \rho+\ldots\Big),\right.\labell{fiexa1}
\eeqa
the logarithmic term is involved only when $d-\alpha$ is even. For the specific value of $\alpha$ given here, it is concluded that we have the logarithmic term at order $\rho$.

To obtain the general form for the anomaly in the presence of scalar field, as a covariant expression of terms with two derivatives, one requires to find only the order $\al+1$ in the expansion of the square root which contributes in the anomaly, (see \reef{resclag}). The result for the expansion of the square root of the determinant of the metric \reef{gex00} by using \reef{detex}, that includes the power $\al+1$ of $\rho$, is
\beqa
\sqrt{det g}&=&\sqrt{det g_{(0)}}\rho^{\al+1}\bigg(\frac{1}{2}{(g^{-1}_{(0)})}^{k l}{g^{(\al+1)}}_{ k l}-\frac{1}{2}{(g^{-1}_{(0)})}^{k l}{g^{(1)}}_{ j l}{(g^{-1}_{(0)})}^{j m}g^{(\al)}_{ k m}\nonumber\\
&&\left.+\frac{1}{4}{(g^{-1}_{(0)})}^{k l}g^{(1)}_{ k l}{(g^{-1}_{(0)})}^{j m}g^{(\al)}_{ j m}\bigg).\right.\labell{detga1}
\eeqa
The anomaly is obtained by substitution \reef{ga1} and \reef{g1,g2} in the first term in the expression \reef{totalanomaly} and the gravitational contribution to the anomaly is found as
\beqa
\cA_g &=& -2 \frac{C_1 d}{16\pi G_N^{(d+1)}}\bigg(\frac{(2-4 b_1+d (-1+4 b_1+4 (d-1) b_2)) }{8 (d-1)}R {(\p^{(0)})}^2\nonumber\\
&&\left.+\frac{1}{2} (b_4 + d\ b_5)\square{(\p^{(0)})}^2+\frac{1}{2} (b_3 + d\ b_6)\nabla_k\p^{(0)}\nabla^k \p^{(0)}\bigg),\right.\labell{anomaly}
\eeqa
where $\Box$ indicates the covariant derivatives with respect to $g^{(0)}$.
The second term is a total derivative and is scheme dependent, i.e. it can be removed by adding a covariant counterterm to the on-shell action \cite{Haro:2001hr}. The overall factor is ascertained by solving equations of motion directly and is presented in \reef{ATa1}. Via substitution values b's in \reef{b1b4}, the ratio of non vanishing terms is
\beqa
\cA_g &=& C(\frac{d-2}{4(d-1)}R {(\p^{(0)})}^2+\nabla_k\p^{(0)}\nabla^k \p^{(0)}),\labell{anomalyratio}
\eeqa
where $C$ is given by
\beqa
C=-\ \frac{C_1 d}{16\pi G_N^{(d+1)}} (b_3 + d\ b_6).\nonumber
\eeqa
The formula \reef{anomalyratio} is in exact agreement with the result obtained in \cite{graham:2001} for the covariant operator with two derivatives apart from a minus sign due to our convention for the definition of the curvature tensor. However, this result only determines the covariant form of the anomaly since from the equations of motion $b_5$ and $b_6$ become singular for $\alpha=\frac{d}{2}-1$.

The matter contribution to the anomaly, coming from the second term in \reef{totalanomaly}, involves $\phi^{(1)}$ that can be obtained in the similar fashion by the PBH transformations. The following expression satisfies the transformation given in \reef{fi012}
\beqa
\phi^{(1)}=\frac{L^2}{2(-2\alpha-2+d)}\Big(\frac{\alpha}{2(1-d)}R  \p^{(0)} +\Box \p^{(0)}\Big).\labell{pbhfi1}
\eeqa
The ratio of two coefficients in the brackets for $\alpha=\frac{d}{2}-1$ and after using by part integration to the last terms is the same as the gravitational conformal anomaly in \reef{anomalyratio}. However, the above expression is again singular at order which $\rho^{k}=\rho$. It means the PBH approach is only useful in order to determine the covariant form of the anomaly although it fails to determine the overall factor. We come back to this issue in more details in section \ref{EOMdk1}.

Using the fact that the anomaly is conformally covariant \cite{Deser:1993gc}, \ie it transforms up to a total derivative, we check that the variation of the expression \reef{detga1} under the PBH transformations must be at most a total derivative. For this purpose, first notice that only the variation of $g^{(1)}$ and $g^{(\al+1)}$ are required to be considered while the variation of other elements only produces homogenous terms. The first term in the PBH transformation of $g^{(\al+1)}$ in \reef{dga12} is homogenous, also since the covariant derivative of $g_{(0)}$ vanishes the last term is a total derivative, and we do not consider it as it is scheme dependence. After substituting from \reef{abeta} and taking trace, we find
\beqa
\Tr(\del g^{(\al+1)}_{i j}(x))=\frac{C_1}{2} L^2 (-\ \frac{d}{2}+1)\Box \sig {(\p^{(0)})}^2+\textrm{Homogenous term}+\textrm{Total derivative}.\nonumber\\\labell{trdga1}\eeqa
Similarly, from the transformation of $g_{(1)}$ in \reef{dga12}, we have
\beqa
\Tr(\del g^{(1)}_{i j}(x))=\Tr(2\partial_{(i} a^k_{(1)} g^{(\al)}_{j) k}(x))= L^2 \Box \sig .\labell{trdg1}
\eeqa
From \reef{ga1} the variation of $g^{(\alpha)}$ is homogenous we only need to consider the transformations
\reef{trdga1} and \reef{trdg1} in \reef{detga1} that manifests the variation of the trace anomaly vanishes under the PBH transformations.
\subsection{k=2, general d} \label{k=2pbh}
In this case to procure the anomaly for $k=2$, $\ie$ four derivatives, it is necessitated considering the order $\al+2$ in the expansion of the square root in \reef{detex}. It implies to take terms terms at order $\al+2$ in \reef{resclag}. The explicit expansion of the metric in this case is
\beqa
g(x,\rho)&\!\!\!\!=\!\!\!\!
&g^{(0)}{(x)}+g^{(1)}{(x)}\rho+\ldots+g^{(n)}{(x)}\rho^n\nonumber\\
&&\left.+\rho^{\al}\bigg(g^{(\al)}{(x)}
+g^{(\al+1)}{(x)}\rho+g^{(\al+2)}{(x)}\rho^2+h^{(\al+2)}{(x)}\rho^2
log \rho+\ldots\bigg).\right.\labell{gexa2}
\eeqa
To acquire  $g^{(\al+2)}$, it is essential to take into consideration all possible covariant terms with the conformal dimension $2(\al+1)$ and with two free indices; As a result the feasible contributions come from the terms with two scalar fields having the conformal dimension $\al$ and four derivatives which is equal to the number of derivatives that emerge in the variation of \reef{gexa2}. Then by applying the BPH transformations to the covariant expression and comparison with the last term in \reef{dga12}, one could find the unknown coefficients in this combination.
However, due to the appearance of the trace of  $g^{(\al+2)}$ in the anomaly which is achieved from the equations \reef{greom}, we do not present the explicit form of it here. Refereing to \reef{detex} and \reef{B} at order $\al+2$, for finding the formula for the gravitational conformal anomaly with four derivatives, we use the following expression:
\beqa
\sqrt{det g}&=&\sqrt{det g_{(0)}}\rho^{\al+2}\bigg(\frac{1}{2}\Tr(g^{-1}_{(0)}g^{(\al+2)})-\frac{2}{4}\Tr(g^{-1}_{(0)} g^{(1)}g^{-1}_{(0)}g^{(\al+1)})\nonumber\\
&&\left.-\frac{2}{4}\Tr(g^{-1}_{(0)} g^{(2)}g^{-1}_{(0)}g^{(\al)})+\frac{2}{8}\Tr(g^{-1}_{(0)}g^{(1)})\Tr(g^{-1}_{(0)}g^{(\al+1)})
\right.\nonumber\\
&&\left.+\frac{2}{8}\Tr(g^{-1}_{(0)}g^{(2)})\Tr(g^{-1}_{(0)}g^{(\al)})-\frac{1}{8}\Tr(g^{-1}_{(0)}g^{(1)}
g^{-1}_{(0)}g^{(1)})\Tr(g^{-1}_{(0)}g^{(\al)})\right.\nonumber\\
&&\left.-\frac{2}{8}\Tr(g^{-1}_{(0)}g^{(1)}g^{-1}_{(0)}g^{(\al)})\Tr(g^{-1}_{(0)}g^{(1)})
+\frac{3}{6}\Tr(g^{-1}_{(0)}g^{(1)}g^{-1}_{(0)}g^{(1)}g^{-1}_{(0)}g^{(\al)})\right.\nonumber\\
&&\left.+\frac{3}{48}\Tr(g^{-1}_{(0)}g^{(1)})\Tr(g^{-1}_{(0)}g^{(1)})\Tr(g^{-1}_{(0)}g^{(\al)})\bigg).\right.\labell{detgal+2}
\eeqa
In the previous section, we saw the conformally invariant form of the scalar field at order that the logarithmic term exists, is the same as one for the conformal anomaly built of the scalar field. We use this fact to find the covariant form for total anomaly.

In the considering case, $\phi^{(2)}$ appears in the matter part for computing the anomaly. To obtain the form of it by applying the PBH transformations, we write down the most general covariant expression consist of four derivatives and one scalar field with undetermined coefficients which are identified by comparing the PBH transformation of $\phi^{(2)}$ with \reef{fi012}.
\beqa
\phi^{(2)}_{PBH}&\!\!\!\!=\!\!\!\!&p_1  C_{ k l m n}C^{ k l m n}\phi^{(0)}+L^4\bigg( \Box\Box\phi^{(0)}-\frac{ \alpha }{2 (d-1) }\square R\phi^{(0)}
-\frac{\alpha (-2+d-2 \alpha )}{ (d-2)^2 } R_{i j} R^{i j}\nonumber\\&&\left.\phi^{(0)}+\frac{ \alpha  \left(d^2 (5+\alpha )+4 (4+3 \alpha )-2 d (9+5 \alpha )\right)}{4 (d-2)^2 (d-1)^2 } R^2\phi^{(0)}-\frac{ (-4 (1+\alpha )+d (2+\alpha ))}{ (d-2) (d-1) }\right.\nonumber\\&&\left. R \square\phi^{(0)}+\frac{ (-2+d-4 \alpha )}{2 (d-1) }\partial_i R \partial^i \phi^{(0)}
+\frac{2 (-2+d-2 \alpha )}{ (d-2) } R_{i j} \nabla^i\nabla^j \phi^{(0)}
\bigg)\right.\nonumber\\&&\left./\bigg(8 (-4+d-2 \alpha ) (-2+d-2 \alpha )\bigg),\right.\labell{fi2pbh}
\eeqa
where $\al$ is the conformal dimension of $\phi^{(0)}$ and the first term involving the Weyl tensor transforms to itself apart from a constant factor.
We will see in \reef{fi2p} that the Weyl tensor is not required in \reef{fi2pbh}, however, as commented in the discussion section, in general there is feasibility to induct it. We expect that above expression is also proportional to the corresponding anomaly \reef{totalanomaly} (see section \ref{EOMdk2}).

In fact if we use the Panetiz operator with four derivatives introduced in \cite{graham:2001} acting on one scalar field we have the following expression that from the property of this operator is covariant under the PBH transformations
\beqa
 P_2 \phi^{(0)}&\sim&\Box \Box \phi^{(0)}-(d-4)\bigg(\frac{1}{4(d-1)}\Box R \phi^{(0)}+\frac{1}{(d-2)^2}R_{i j} R^{i j}\phi^{(0)}\bigg)-R^2 \phi^{(0)}\nonumber\\
&&\left.\times\frac{(d-4) }{4 (d-1)^2}\bigg(\frac{4-3 d}{(d-2)^2}-\frac{d}{4}\bigg)-\frac{(8+(d-4) d) }{2 (d-2) (d-1)}R \Box \phi^{(0)}+\frac{4\ R_{i j}\partial^i \partial^j\phi^{(0)} }{d-2}\right.\nonumber\\
&&\left.-\ \frac{(d-6)}{2 (d-1)}\partial_i R \partial^i \phi^{(0)},\right.\labell{math4}
\eeqa
one can see that is proportional to the formula \reef{fi2pbh} for $\alpha=\frac{d}{2}-2$.

Since the anomaly is a covariant object under the transformations, we expect \reef{math4} is actually part of the anomaly that includes terms with two scalars and four derivatives.
However, it is not the most practicable covariant form. For instance a combination in terms of Weyl tensor and/or the terms with more than two scalar fields may be added, depending on the form of the action (see the discussion section for further illustration).

To show above expression is covariant under transformations given in the formulas \reef{dga12}, \reef{Rtrans} and \reef{fi012} we use the fact that in general the variation of an expression with any number of covariant derivatives and tensor indices is
\beqa
\delta [\nabla_{k_1}\nabla_{k_2}\ldots \nabla_{k_n}A_{i_1 i_2\ldots i_m}]&=&-\delta \Gamma_{k_1 k_2}^{k}\nabla_{k}\nabla_{k_3}\ldots \nabla_{k_n}A_{i_1 i_2\ldots i_m}-\ldots\nonumber\\
&&\left.-\delta \Gamma_{k_1 i_1}^{k}\nabla_{k_2}\nabla_{k_3}\ldots \nabla_{k_n}A_{k i_2\ldots i_m}-\ldots\right.\nonumber\\
&&\left.-\nabla_{k_1}[\delta \Gamma_{k_2 k_3}^{k}\nabla_{k}\nabla_{k_4}\ldots \nabla_{k_n}A_{i_1 i_2\ldots i_m}+\ldots]\right.\nonumber\\
&&\left.-\nabla_{k_1}\nabla_{k_2}\ldots \nabla_{k_{n-1}}[\delta \Gamma_{k_n i_1}^{k}A_{k i_2\ldots i_m}+\ldots]\right.\nonumber\\
&&\left.+\nabla_{k_1}\nabla_{k_2}\ldots \nabla_{k_{n}}[\delta A_{i_1 i_2\ldots i_m}],\right.\labell{multi deriv}
\eeqa
where the variation of the Christoffel symbol is given by
\beqa
\delta \Gamma_{i j}^{k}=\partial_i\sigma\delta_j^k+\partial_j\sigma\delta_i^k-\partial^k\sigma g^{(0)}_{i j}.\labell{gamma var}
\eeqa
To corroborate that \reef{math4}is covariant under the transformation, since the transformation of the inverse of the boundary metric and the boundary value of the scalar field is homogeneous therefore
we simply do not need to consider them. For instance more with regarding \reef{multi deriv} and \reef{gamma var} for a scalar field $\phi$ we have
\beqa
\delta \Box\phi\rightarrow\Box \delta \phi-(2-d)\nabla_i \sigma\nabla^i \phi .\nonumber
\eeqa
However, When we encounter the variation of derivatives acting on a tensor, one must consider contributions of transformations of the homogenous parts or the tensor, in particular the homogenous part of transformation for the Ricci scalar \reef{Rtrans}.

Nonetheless it is realized  that the PBH transformations do not give us all participation in the anomaly such as contributions with different number of scalars as we encounter in section \ref{n2k1}. Furthermore these transformations may not fix all the coefficients in the covariant expression. We demonstrate the formula for the anomaly for specific and general dimensions through solving the equations of motion in the next section, that is, we reproduce all the coefficients of the covariant expression for the anomaly that was given by the PBH transformations as well as the overall factor.

\section{Anomaly through solving the back reacted gravitational equations of motion} \label{EOM}
The conformal dimension of the dual operator is determined by considering that the following expression is covariant under the conformal transformations \cite{graham:2001}
\beqa
\int \sqrt{g} \phi^n P_k  \phi^n .\nonumber
\eeqa
For instance, for $k=1$,
\beqa
P_1=-\Box+\frac{d-2}{4(d-1)}R,
\eeqa
and $P_2$ after acting on a scalar was given in \reef{math4}.\\
Where $P_k$ and n scalar fields transform as \footnote{ Remember that through turning to account conformal invariance to the action we obtain specific values of $\al$ to be $ \frac{d}{2}-k$ where $k$ indicates the number of square derivatives.}
\beqa
P_k\rightarrow e^{(-\frac{n}{2}-k)\sigma}P_k e^{(\frac{n}{2}-k)\sigma},\qquad \qquad \phi^n\rightarrow e^{-\alpha n \sigma}\phi^n,
\eeqa
and by considering the transformation $ \sqrt{g}$, \ie $e^{d \sigma }$, it turns out that for the dimension d and 2n scalar fields, one has
\beqa
\alpha&=&\frac{d-2k}{2 n}.\labell{alfa}
\eeqa
Therefore, the result for the anomaly for a general dimension is presented in different categories that are characterized by the choice of the number of scalar fields 2n and of derivatives k. However, the anomaly formula for a specific dimension may involve all possible number of scalar fields and their derivatives as well as the Ricci tensor such that each individual term has the desired conformal dimensions.
In this section, first we follow the procedure given in section \ref{asym} to find the conformal anomaly in the case of $n=1$, first for the specific spacetime dimensions $d=4,6$ and the number of derivatives  specified by $k=1,2$, we generalize this approach to achieve the anomaly for a general dimension and $k=1,2$. Then in the last three sections we explore the result in the case of $n=2$.
\subsection{k=2, d=4} \label{EOM4}
For a dynamical AdS background, \ie the gravity coupled to scalar, the gravitational field equations are given in \reef{greom}. The approach is the same as the pure gravity, it means we solve the equations of motion by applying the CFT data.
For illustration, we reveal the formula of the conformal anomaly for $d=4$ and the massless scalar field, therefore $\Delta=4$ and $\alpha=0$. The explicit form of the expansions of the scalar field and metric that construct the anomaly with four derivatives and with two scalar fields only involves integer powers of $\rho$ that are as follows:
\beqa
\phi(x,\rho)&=&\phi^{(0)}+\rho \phi^{(1)}+\rho^2 \phi^{(2)}+\rho^2\ \log\rho \psi^{(2)},\nonumber\\
{g}(x,\rho)&=&{g^{(0)}}+\rho {g^{(1)}}+\rho^2 {g^{(2)}}+\rho^2\ \log\rho {h^{(2)}}.\labell{exd2}
\eeqa
The recursion relation for $g^{(1)}$ is determined by zero order of the perturbative expansion of the first equation in \reef{greom}
\beqa
g^{(1)}_{i j}=\frac{1}{d-2}\left(- \Tr(g^{(1)})g_{i j}^{(0)}-R_{i j}+8 \pi G_N   \partial_i\phi^{(0)}\partial_j\phi^{(0)}\right),\labell{g1d4}
\eeqa
where
\beqa
\Tr g^{(1)}=\frac{-R+8 \pi G_N   \partial_k\phi^{(0)}\partial^k\phi^{(0)} }{2 (d-1)}.\labell{TRg1d4}
\eeqa
From the third equation in the leading order, one gets
\beqa
\Tr g^{(2)}=\frac{1}{4}\Tr(g_{(0)}^{(-1)}g^{(1)}g_{(0)}^{(-1)}g^{(1)})-8 \pi G_N (\phi^{(1)})^2.\labell{trg2d4}
\eeqa
Also trace of ${h^{(2)}}$ from  $\rho \log\rho$ part of the expansion is zero.
Heed that for the massless scalar field because $m^2$ is zero only the gravitational constituent of the anomaly in \reef{totalanomaly} prevails. The anomaly is achievable  from \reef{detex} in which only the first two terms of $B$ exist in \reef{B}\footnote{From now on we suppress factors of $g_{(0)}$. Unless we mean the contrary, indices are raised and lowered with metric $g_{(0)}$ also all contractions are built with this metric.}
\beqa
\cA^{(\alpha=0)}&=&- \frac{2d}{16 \pi G_N}\Big(\frac{1}{2}\Tr g^{(2)}  - \frac{1}{4}\Tr(g^{(1)}g^{(1)})+ \frac{1}{8}(\Tr {g^{(1)}})^2\Big),\labell{Ad4}
\eeqa
where we have used the following expression from the solution of the scalar equation \reef{seq} that is linear in the scalar field at the leading order
\beqa
\phi^{(1)}=\frac{1}{4}\square\phi^{(0)},
\eeqa
and $g^{(1)}$ is defined according to \reef{g1d4}.

The second equation in \reef{greom} provides nonlinear relations between different components of the metric and the scalar field, \eg  the trace of $g^{(1)}$ gives a relation between its divergence and $\phi^{(1)}$ as
\beqa
\nabla_i(Tr g^{(1)})-\nabla^j g^{(1)}_{i j}=-16 \pi G_N \phi^{(1)}\partial_i\phi^{(1)}.\labell{dif eq}
\eeqa
In conclusion, the formula for the anomaly after substitution \reef{trg2d4} in \reef{Ad4} is
\beqa
\cA^{(\alpha=0)}&\!\!\!\!=\!\!\!\!
&- \frac{-2}{384 G \pi }\Big(-R^2+3 R^{i j}R_{i j}+2 (8\pi G_N \partial_i\phi^{(0)} \partial^i\phi^{(0)})^2\nonumber\\
&&\left.+8 \pi G_N \big(-6 R_{i j}\partial^i\phi^{(0)}\partial^j\phi^{(0)}+2\ R \ \partial_i\phi^{(0)}\partial^i \phi^{(0)}+3\ \square (\phi^{(0)} )^2\big)\Big),\right.\labell{AD4}
\eeqa
where one can see the pure gravitational part has the same ratio of coefficients given in \cite{Henningson:1998wel}. In other words, when $\phi^{(0)}$ is constant, our result for anomaly is the same as the conclusion in \cite{Henningson:1998wel} for the coefficient of the $\log$ in the on-shell action
\beqa
a_{(2)}=-\ \frac{1}{8\pi G_N}(E_{(4)}+I_{(4)}),\nonumber
\eeqa
where $E_{(4)}$ and $I_{(4)}$ are 4 dimensional Euler density and a conformal invariant, respectively that are introduced in \cite{Henningson:1998wel}.
For the rest of terms, it is easy to see that the above formula for the anomaly turns into \reef{math4} by virtue of Bianchi identity and suppressing total derivatives by choice of the scheme through adding  local finite counterterms, that is implied the following relations:
\beqa
R_{i j} \partial^i \phi^{(0)}\partial^j\phi^{(0)}&\rightarrow&\frac{1}{4}R \square (\phi^{(0)})^2-R_{i j}  \phi^{(0)}\nabla^i\nabla^j\phi^{(0)},\labell{bia1}\\
R \partial_i \phi^{(0)}\partial^i\phi^{(0)}&\rightarrow&\frac{1}{2}R \Box (\phi^{(0)})^2-R \phi^{(0)}\square \phi^{(0)}.\nonumber
\eeqa
Also the anomaly receives a contribution of four scalar fields with derivatives anting on in the last term of the first line in \reef{AD4} that is covariant under the transformations, since in the four dimensional spacetime, the scalar field does not transform. Therefore, the formula \reef{math4} should be modified for specific values of dimension. Though it is matched with the remaining terms in \reef{AD4}, in particular, it does not involve $R^2 (\phi_{(0)})^2$.
 Note that we only get the terms with derivatives acting on the scalars and not some power of the scalar field.
\subsection{ k=1, d=6} \label{EOM6}
In this section, we express an obstacle for finding the anomaly and present a way to handle it. In general the form of the expansion of the metric depends on dimension, however, for the scalar field for a specific $k$, for any dimension, it remains unchanged. In the six dimensional spacetime, the expansions of the metric and the scalar field are as follows:
\beqa
{g}(x,\rho)&=&{g^{(0)}}+\rho\ {g^{(1)}}+\rho^2 {g^{(2)}}+\rho^3 {g^{(3)}}+\rho^3\ \log\rho {h^{(3)}}+\ldots ,\nonumber\\
\phi(x,\rho)&=&\phi^{(0)}+\rho\ \phi^{(1)}+\rho\ \log\rho \psi^{(1)}+\rho^2 \phi^{(2)}+\ldots .\nonumber
\eeqa
In this case, by considering the first equation at the leading order we get the same result for $g^{(1)}$ in \reef{g1,g2}. However, the expression for $g^{(2)}$ resulting from the first order, must be modified as follows for $d=6$ and $\alpha=2$:
\beqa
g^{(2)}_{i j}&\!\!\!\!=\!\!\!\!&-\ \frac{1}{2 (-4+d) }\bigg( R^{(1)}_{i j}+2 g^{(1)}_{i k}g^{(1)k}_{j}- \big(\Tr( g^{(1)}g^{(1)})-2 \Tr g^{(2)}\big)g^{(0)}_{i j}\nonumber\\
&&\left.
+\ \frac{16 (d-2) G \pi }{d-1} {\phi_{(0)}}^2g^{(0)}_{i j}\bigg),\right.\labell{g2d6}
\eeqa
where $R^{(1)}_{i j}$ is the first order term in the expansion of the Ricci tensor
\beqa
R^{(1)}_{i j}&\!\!\!\!=\!\!\!\!
&-\ \frac{1}{2}\square g^{(1)}_{i j}+\frac{1}{2}\partial_k\partial_i g^{(1)}_{j k}+\frac{1}{2}\partial_k\partial_j g^{(1)}_{i k}-\frac{1}{2}\partial_i\partial_j \Tr g^{(1)}\nonumber\\
&&\left.
=\frac{L^2}{4(d-2)(d-1)}\Big(2(d-1)\partial_k \partial^k R_{i j}-(d-2)\partial_i \partial_j R+4(d-1)(R_{i k}{R_{ j}}^{k}\right.\nonumber\\
&&\left.-R^{k l}R_{l i k j})-\partial_k \partial^k R g^{(0)}_{i j}\Big).\right.\labell{Ric1}
\eeqa
In fact, one can see the result in \reef{g2d6} apart from the last term, by virtue Bianchi identity, is the same as the second formula in \reef{g1,g2}. The Ricci scalar is zero at this order. Also $\Tr g^{(2)}$ that is another element for constructing of the anomaly is procured directly from the third equation at the leading order as
\beqa
\Tr g^{(2)}=\frac{1}{4}\Tr( g^{(1)}g^{(1)})-\frac{4 \pi d G_N }{(d-1)}{\phi_{(0)}}^2.\labell{Trg2}
\eeqa
By referring to the formula \reef{detex} one needs to find $\Tr g^{(3)}$ as well that is achieved from the third equation at order $\rho$ as
\beqa
\Tr g^{(3)}&\!\!\!\!=\!\!\!\!& \frac{1}{6} \Big(-\Tr(g^{(1)}g^{(1)} g^{(1)})+4 \Tr(g^{(1)}g^{(2)})-5 \Tr h^{(3)}+8  \pi G_N\big(-4\phi_{(0)} \psi_{(1)}\nonumber\\
&&\left.+\frac{2 (2-3 d) }{d-1}\phi_{(0)}\phi_{(1)} \big) \Big),\right.\labell{trg3}
\eeqa
however, the explicit form of $g^{(3)}$ is not granted by means of solving equations of motion.

In the absence of the scalar field, $h^{(\frac{d}{2})}$ is traceless and covariantly conserved. However, the existence of relevant deformations at the same order, gives us
\beqa
\Tr h^{(3)}&=& -\ \frac{ (-2+3 d) 8\pi G_N  }{3 (d-1)}\phi_{(0)}\psi_{(1)}.\labell{trh3}
\eeqa
The form of $\psi^{(1)}$ is acquired by taking the scalar equation into consideration
\beqa
\psi^{(1)}&=&-\ \frac{1}{4}\Big( \alpha \Tr g^{(1)}  \phi^{(0)}+\Box \phi^{(0)}\Big).\labell{psi1}
\eeqa
According to \reef{detex}, the formula for the gravitational part of the anomaly is given by
\beqa
\cA_g^{(\alpha=2)}&\!\!\!\!=\!\!\!\!&-\ \frac{6 }{16 \pi G_N}\bigg(\frac{1}{2}\Tr g^{(3)} -\frac{1}{2}\Tr(g^{(1)}g^{(2)}) +\frac{1}{4}\Tr g^{(1)} \Tr g^{(2)} -\frac{1}{8} \Tr g^{(1)}  \Tr(g^{(1)}g^{(1)})\nonumber\\
&&\left.+\frac{1}{6} \Tr(g^{(1)}g^{(1)}g^{(1)})+\frac{1}{48}\big(\Tr g^{(1)} \big)^3 \bigg).\right.
\eeqa
Now after exerting \reef{trg3} with usage relations \reef{g2d6} and \reef{Trg2}, we obtain
\beqa
\cA_g^{(\alpha=2)}&\!\!\!\!=\!\!\!\!&\frac{-2d}{768 (d-4) (d-2) (d-1)  \pi G_N} \bigg(-\Box R\ \Tr g^{(1)}+(d-2) (d-1)\Big((d-4) \nonumber\\
&&\left.(\Tr g^{(1)})^3+(10-3 d) \Tr g^{(1)}\Tr(g^{(1)}g^{(1)}) +4 (d-4)\Tr(g^{(1)}g^{(1)}g^{(1)}) \Big)+ g_{(1)}^{i j} \right.\nonumber\\
&&\left. \Big(2 (d-1) \Box R_{i j}-(d-2) \partial_i \partial_j R+4 (d-1) \big(2 (d-2) {g^{(1)}}_{i}^{k}g^{(1)}_{k j}-R_{i}^{k} R_{k j}\right.\nonumber\\
&&\left. +R_{k i l j}R^{k l}\big)\Big)\bigg)-\frac{5 d }{192  \pi G_N}\Tr h^{(3)}-\frac{d}{6}  \phi^{(0)}\psi^{(1)}-\frac{d (-2+3 d) }{48 (d-1)} \Big(\Tr g^{(1)}(\phi^{(0)})^2 \right.\nonumber\\
&&\left.+4 \phi^{(0)}\phi^{(1)}\Big).\right.\labell{A1d6}
\eeqa
However, the relation \reef{A1d6} is not a combination resembling \reef{anomalyratio} at the part involving the scalar field. In particular, the explicit form of $\phi^{(1)}$ is not determined from the equations because the coefficient of this term at the first order of the scalar equation \reef{seq} vanishes and at the next, order there is a relation between $\Box \phi^{(1)}$ and higher order field, $\phi^{(2)}$. In other words, $\phi^{(1)}$ as a nonlocal field is not ascertained via information at the boundary, \ie $g^{(0)}$ and $\phi^{(0)}$. However, there is another contribution to the anomaly that comes from the mass term in the action \cite{Haro:2001hr} that was not involved in the previous section due to the massless scalar.
In the expansion of $\sqrt{g}\ ({\phi} (x,\rho))^2$ in \reef{resclag}, only the terms at order one take part in the matter anomaly, and the result is
\beqa
\cA_{M}&=&\frac{-2\al(d-\al)}{2(d-1)}\big(2 \phi_{(0)}\phi_{(1)}+\frac{1}{2}\Tr(g^{(1)})(\phi_{(0)})^2\big).\labell{AM}
\eeqa
In fact, the total anomaly is given by adding \reef{A1d6} and \reef{AM} involve terms with scalars that are proportional to $\psi^{(1)}$ which is covariant under transformations, so the covariance under transformations is recovered. The total anomaly for $d=6$ is given by
\beqa
\cA_t^{(\alpha=2)}&\!\!\!\!=\!\!\!\!&\frac{-2}{6\times 2^{10} (d-4) (d-1)^2 (d-2)^2  \pi G_N }\bigg(4 d\ (d-1) \Big( d\ R\  \Box R +(2+3 d) R^{i j}R_{i j}\nonumber\\
&&\left.R-4(d-1) \big(R^{i j} \Box R_{i j}+2R^{i j} R^{k l} R_{i k j l}\big)\Big)- (2+d) d^2 R^3\bigg)-\ \frac{1}{288 (d-1)^2}\right.\nonumber\\
&&\left.\times\Big(-4 \big(36+d (-19+3 d)\big) R (\phi^{(0)})^2+d(d-1)(2+3 d) \phi^{(0)}\Box\phi^{(0)}\Big).\right.\labell{ATd6}
\eeqa
The first two lines reproduce the results given in \cite{Henningson:1998wel}\footnote{beside a minus sign in our convention for the definition of the Ricci tensor and by virtue Bianchi identity to the first term.} in the absence of the scalar field. Also the ratio of terms with two scalars is exactly the same as what we expect in \reef{anomalyratio} for the six dimensional spacetime.
\subsection{k=1, general d} \label{EOMdk1}
We now proceed to obtain the solution for a general dimension while the relation \reef{alfa} is satisfied, it means we set a constraint on the mass of the scalar field, on other words, for other values of the mass basically there is no anomaly involving the scalar field, therefore we are only interested in the solution in which the anomaly appears. In order to find the anomaly in the presence of scalar field for a general dimension and of only terms with two derivatives, we follow the procedure similar to the previous section. The formula given in \reef{ga1} by using the PBH transformations is singular for $\alpha=\frac{d}{2}-1$, so we employ the equations of motion to attain the anomaly. Here we suppose that $\alpha>1$ (or $d>4$), therefore the formula for $g^{(1)}$ given in \reef{g1,g2} is applicable to this case.\footnote{Although in the next section we refer to the formula in this section for $\alpha>2$.}
Via making approach to the formula \reef{detex}, one demands to solve the third equation at order $\alpha-1$ to find
\beqa
\Tr g^{(\alpha+1)}&=&-\ \frac{1}{\alpha(1+\alpha)(d-1)}\bigg((d-1)\Big((1+2\alpha)\Tr h^{(\alpha+1)}-C_1\alpha^2 \Tr g^{(1)}  (\phi^{(0)})^2\Big) \nonumber\\
&&\left.+8 \pi G^{d+1}_N \alpha \Big((-2+d+d \alpha)\phi^{(0)}\phi^{(1)}+2(d-1)\phi^{(0)}\psi^{(1)}\Big)\bigg),\right.\labell{TRgal1}
\eeqa
where $C_1$ is obtained by setting \reef{ga1} in the third equation and considering the order $\alpha-2$
\beqa
C_1=-\ \frac{4 \pi G^{d+1}_N}{d-1},\labell{C1}
\eeqa
also the coefficient of the logarithmic term at this order is
\beqa
\Tr h^{(\alpha+1)}&=&-\ \frac{8 \pi G^{d+1}_N (-2+d+d \alpha ) \phi^{(0)}\psi^{(1)} }{(1+\alpha) (d-1)},\labell{TRhal1}
\eeqa
where $\psi^{(1)}$ is the same as before \reef{psi1}. Indeed, this form of $\psi^{(1)}$ is universal and is applicable to any specific dimension with the same number of derivatives and of scalar fields in the anomaly. However, $g^{(\alpha+1)}$ is not obtainable through solving the equations of motion since at this order its coefficient vanishes. We find this component for a different value of $\alpha$ in the next section.

The gravitational part of the anomaly by noticing the formula \reef{detga1} and \reef{TRgal1} is specified as
\beqa
\cA_g^{(\alpha=\frac{d}{2}-1)}&\!\!\!\!=\!\!\!\!&-\ \frac{-2d }{16 \pi G^{d+1}_N}\bigg( \frac{(1+2 \alpha )}{2 \alpha  (1+\alpha )}\Tr h^{(\alpha+1)}+\frac{ \pi G_N  (-2+d+d \alpha )  }{(d-1) (1+\alpha )}\Big(\Tr g^{(1)}  (\phi^{(0)})^2\nonumber\\
&&\left.+4 \phi^{(0)}\phi^{(1)}\Big)
+\frac{8  \pi G^{d+1}_N }{1+\alpha }\phi^{(0)}\psi^{(1)}\bigg),\right.\labell{A1al1}
\eeqa
and the matter anomaly is the same as \reef{AM}. In conclusion, only a combination proportional to  $\psi^{(1)}$ remains in the total anomaly, see \reef{TRhal1}.
After substitution of the trace of $g^{(1)}$
in \reef{g1,g2} and the expressions \reef{TRgal1}, \reef{TRhal1} and removing the total derivative, for $\alpha=\frac{d}{2}-1$, we attain
\beqa
\cA_t^{(\alpha=\frac{d}{2}-1)}&\!\!\!\!=\!\!\!\!&\frac{-2}{2d}\bigg(\frac{(d-2) }{4 (d-1)}R (\phi^{(0)})^2+\partial^k\phi^{(0)} \partial_k\phi^{(0)}\bigg),\labell{ATa1}
\eeqa
and therefore the anomaly maintains the covariant description in \reef{anomalyratio}.
\subsection{k=2, general d} \label{EOMdk2}
In this section, we aim to find the anomaly which has the structure of four derivatives and of two scalar fields, for a general dimension, through solving the equations of motion. For this purpose similar to before we solve the coupled system of the first and third equations in \reef{greom} by inserting the expansions \reef{gexa2} and the first term in \reef{exd2} into the equations that is straightforward but somewhat tedious. Again note that this progress breaks down at $\alpha=\frac{d}{2}-2$ and the coefficient of $\phi ^{2}$ vanishes, therefore a logarithmic term is required to get the full solution of the back-reaction of the scalars to the bulk geometry. From the formula \reef{detgal+2}, we need to pursue different components in order to construct the anomaly. The explicit form of $g^{(\alpha+1)}$ is determined via the first equation in \reef{greom} at order $\alpha$
\beqa
g^{(\alpha+1)}_{i j}&\!\!\!\!=\!\!\!\!&\frac{1}{(d-1) (1+\alpha ) (2-d+2 \alpha )}\bigg((d-1) \Big(R^{(\alpha)}_{i j} +(1+\alpha )\Tr g^{({\alpha +1})} g^{(0)}_{i j} \Big)
 \nonumber\\
&&\left.+8  \pi G^{d+1}_N\Big((-2+d-\alpha ) \alpha g^{(1)}_{i j}  (\phi^{(0)})^2+\alpha  g^{(0)}_{i j}  \big(\Tr g^{(1)}  (\phi^{(0)})^2+2 (d-\alpha ) \phi^{(0)}\phi^{(1)}\big)\right.\nonumber\\
&&\left.
-(d-1) \partial_i \phi^{(0)} \partial_j\phi^{(0)}\Big)\bigg),\right.\labell{Gad1}
\eeqa
where the expansion of the Ricci tensor at this order is
\beqa
R^{(\alpha)}_{i j}&\!\!\!\!=\!\!\!\!
&-\frac{1}{2}\partial_k \partial^k g^{(\alpha)}_{i j}+\frac{1}{2}\partial_k\partial_i g^{(\alpha)}_{j k}+\frac{1}{2}\partial_k\partial_j g^{(\alpha)}_{i k}-\frac{1}{2}\partial_i\partial_j \Tr g^{(\alpha)}\nonumber\\
&&\left.
=\frac{C_1}{2}\Big(- g^{(0)}_{i j} \partial_k \partial^k (\phi^{(0)})^2 +(2-d)\partial_i \partial_j (\phi^{(0)})^2\Big),\right.\labell{Rical}
\eeqa
 where it is related to the second derivative of the scalar fields, as expected because only $ g^{(\alpha)}$ gets contribution at this order. The Ricci scalar reads
\beqa
R^{(\alpha)}=(1-d)C_1 \partial_k \partial^k (\phi^{(0)})^2.\labell{Ral}
\eeqa
The trace of $g^{(\alpha+1)}$ is the same as \reef{TRgal1} if we remove the first and last terms. At the same order of the third equation, for given $\alpha$, by considering the cyclic property of the trace, we have a factor of two for the terms including four components of the metric, as a result, we have
\beqa
\Tr g^{(\alpha+2)} &\!\!\!\!=\!\!\!\!& \frac{1}{(d-1) (1+\alpha ) (2+\alpha )}\bigg((d-1) \Big(\Tr( g^{(1)} g^{({\alpha +1})}) (1+\alpha )^2-\Tr h^{({\alpha +2})}  (3+2 \alpha )\Big)\nonumber\\
&&\left.+4  \pi G^{d+1}_N \Big((1+\alpha +\alpha ^2)\Tr(g^{(1)}g^{(1)}) (\phi^{(0)})^2- (2+\alpha +\alpha ^2)\Tr g^{(2)}(\phi^{(0)})^2 \right.\nonumber\\
&&\left.+4 (1+\alpha )-d \big(4+\alpha  (3+\alpha )\big)\Big) (\phi^{(1)})^2-2\alpha \Big(\big(-4+d (3+\alpha )\big)\phi^{(0)}\phi^{(2)}\right.\nonumber\\
&&\left.
+2 (d-1)\phi^{(0)}\psi^{(2)} \Big)\Big)\bigg),\right.\labell{TRgal2}
\eeqa
where we have inserted the components of the inverse of the metric, for instance, we have
\beqa
{G_{(\alpha)}^{(-1) i j}}&=&- g_{(0)}^{i k}g^{(\alpha)}_{k l}g_{(0)}^{l j}.\labell{GALPHA2}
\eeqa
In general the explicit forms of $g^{(1)}_{i j}$ and $g^{(2)}_{i j}$ that are given by the first equation in \reef{greom}, depending on the value of $\al$, can receive extra terms coming from the first term in rhs of this equation, however, for $\al>2$ the formulas resulted from the PBH transformation given in \reef{g1,g2} are applicable.

Similar to before, from the third equation \reef{greom}, the coefficient of $\rho^{\alpha}\log \rho$ indicates
\beqa
\Tr h^{(\alpha+2)}=-\ \frac{8 \pi G^{d+1}_N   \alpha  (-4+3 d+d \alpha ) }{(d-1) (1+\alpha ) (2+\alpha )}\phi^{(0)}\psi^{(2)}.\labell{TRhal2}
\eeqa
Also at order one of the scalar equation \reef{seq}, one gets
\beqa
\phi^{(2)}&\!\!\!\!=\!\!\!\!&\frac{1}{8(d-2 \alpha -4)}\bigg(-2 g_{(1)}^{k l}\partial_k\partial_l\phi^{(0)}  +2\Box\phi^{(1)}+(2\alpha+4) \Tr g^{(1)}\phi^{(1)}+ \partial_k\Tr g^{(1)}\partial^k\phi^{(0)}\nonumber\\
&&\left.-2\partial^k g^{(1)}_{k l} \partial^l \phi^{(0)} -2 \alpha \phi^{(0)}\Big(\Tr(g^{(1)}g^{(1)} )-2\Tr g^{(2)}\Big)\bigg).\right.\labell{fi2p}
\eeqa
Albeit above expression is reaped without considering the logarithmic term and it is not valid for the chosen value of
$\alpha$ in this section, because the expansion at order two is truncated. However, it turns out that this expression is proportional to our result given by the PBH transformations \reef{fi2pbh}

Since $\phi^{(2)}$ for $\alpha=\frac{d}{2}-2$ is singular, it is essential to add a logarithmic term at the corresponding order, and the corresponding coefficient reads as
\beqa
\psi^{(2)}&=&-\ \frac{(d-2 \alpha -4)}{2}\phi^{(2)}.\labell{psiphi}
\eeqa
In fact we will see this covariant combination is also the possible combination that constructs the anomaly.
The explicit form of $\psi^{(2)}$ in terms of curvature and scalar field and its derivatives after substitution of the expressions for $g^{(1)}_{i j}$ and $g^{(2)}_{i j}$ in \reef{psiphi}, is revealed as
\beqa
\psi^{(2)}&\!\!\!\!=\!\!\!\!&-\frac{1}{32} \bigg(  \Box  \Box \phi^{(0)} -\frac{ \big(8+(d-4) d \big)} {2 (d-2) (d-1)}R\ \Box \phi^{(0)}-\frac{ (d-4)}{4 (d-1)}\Box R \phi^{(0)} + { R }^2 \phi^{(0)}\nonumber\\
&&\left.\times\frac{(d-4)  \Big(-16+d  \big(16+(d-4) d \big) \Big) }{16 (d-2)^2 (d-1)^2}-\frac{(d-4)}{(d-2)^2} {R}_{k l}  {R}^{k l} \phi^{(0)}+\frac{4}{d-2} \right.\nonumber\\
&&\left.
{R}_{k l} \partial^k \partial^l\phi^{(0)}-\frac{(d-6)}{2 (d-1)} \partial_k R  \partial^k\phi^{(0)} \bigg)\right.\labell{Psi2}
\eeqa
Notice that $\phi^{(2)}$ was canceled in above expression.

The anomaly pertaining to the gravitational part of the action is given by considering the first part of  \reef{totalanomaly} and the formula \reef{detgal+2}
\beqa
\cA_g^{(\alpha=\frac{d}{2}-2)}&\!\!\!\!=\!\!\!\!&\frac{-2d}{16\pi G^{d+1}_N}\bigg(\frac{1}{2}\Big(\Tr g^{({\alpha +2})} - \Tr(g^{(1)}g^{({\alpha +1})})- \Tr(g^{(2)}g^{({\alpha})})+\Tr(g^{(1)}g^{(1)}g^{({\alpha })})\Big)\nonumber\\
&&\left.+\frac{1}{4}\Big(\Tr g^{(1)} \Tr g^{({\alpha +1})}
+\Tr g^{(2)}  \Tr g^{({\alpha })}- \Tr(g^{(1)}g^{({\alpha })})\Tr g^{(1)}\Big)\right.\nonumber\\
&&\left.-\frac{1}{8} \Tr(g^{(1)}g^{(1)})\Tr g^{({\alpha })} +\frac{3}{48}\big(\Tr g^{(1)}\big)^2 \Tr g^{({\alpha })}\bigg),\right.\nonumber
\eeqa
where we have applied the relation \reef{Gad1}.

The matter anomaly in this case by referring to the second part of \reef{resclag} becomes
\beqa
A_M^{(\alpha=\frac{d}{2}-2)}&\!\!\!\!=\!\!\!\!& \frac{-2\alpha(d-\alpha )  }{2 (d-1)}\bigg( 2\phi^{(0)}\phi^{(2)}+\phi^{(1)}\phi^{(1)}+\frac{1}{2} \Tr g^{(2)}(\phi^{(0)})^2+ \Tr g^{(1)} \phi^{(0)}\phi^{(1)}\nonumber\\
&&\left.-\ \frac{1}{4} \Tr(g^{(1)} g^{(1)} )(\phi^{(0)})^2
+\frac{1}{8} \big(\Tr g^{(1)}\big)^2 (\phi^{(0)})^2\bigg).\right.\nonumber
\eeqa
Adding these two parts gives us the total anomaly
\beqa
\cA_t^{(\alpha=\frac{d}{2}-2)}&\!\!\!\!=\!\!\!\!&- \frac{-2}{8 (d-2)}\bigg(8 (d-4) \phi^{(0)}\psi^{(2)}+\frac{(d-1) }{ \pi G^{d+1}_N} \Tr h^{({\alpha +2})}+(\Box \phi^{(0)})^2+2 (1+\alpha )\nonumber\\
&&\left. \Tr g^{(1)}\phi^{(0)} \Box \phi^{(0)}+ \alpha  (2+\alpha ) \big(\Tr g^{(1)}\big)^2(\phi^{(0)})^2+4 g^{(1)}_{i j}\partial^i \phi^{(0)} \partial^j\phi^{(0)} -\frac{R^{(\alpha)}_{i j}  g_{(1)}^{i j}}{2  \pi G^{d+1}_N }\right.\nonumber\\
&&\left.-\ \frac{(d-2)^2 }{d-1}\Tr(g^{(1)}g^{(1)}) (\phi^{(0)})^2+\frac{4 d }{d-1}\Tr g^{(2)} (\phi^{(0)})^2-\frac{g^{(1)}_{i j}\partial^i \phi^{(0)} \partial^j\phi^{(0)} }{2(d-2)}\bigg).\right.\nonumber\\\labell{ADALd2}
\eeqa
Similar to the previous section the contribution of $\phi^{(2)}$ in the gravitational part of the anomaly, which is not determined through solving the equations of motion, is canceled by the corresponding term in the matter anomaly for $\alpha=d/2-2$.
Since $\Tr h^{(\alpha+2)}$ is proportional to $\psi$ (see \reef{TRhal2}) and is already in the covariant form. We only need to clear that the rest of the terms in above expression are in the same covariant combination.
We call the the collection of terms in the square brackets apart from the first two terms as $\cT$. After substitution different components and using by part integration, we have
\beqa
\cT&\!\!\!\!=\!\!\!\!&- \frac{-2}{8 (d-2)}\bigg((\Box\phi^{(0)})^2-\frac{ (1+\alpha )}{d-1} R \phi^{(0)}\Box \phi^{(0)}
-\frac{(d-4) }{(d-2)^2}R_{i j} R^{i j}(\phi^{(0)})^2 +R^2 (\phi^{(0)})^2\nonumber\\
&&\left.
\frac{ d^2 \big(3+\alpha  (2+\alpha )\big)+4 \big(4+\alpha  (2+\alpha )\big)(1-d) }{4 (d-2)^2 (d-1)^2}
 +\frac{2}{(d-2) (d-1)}R\ \partial^k\phi^{(0)} \partial_k\phi^{(0)} \right.\nonumber\\
&&\left. +\ \frac{1}{d-1}R_{i j}\partial^i \partial^j (\phi^{(0)})^2 -\frac{4 }{d-2}R_{i j}\partial^i \phi^{(0)} \partial^j\phi^{(0)}\bigg),\right.\nonumber
\eeqa
where it is  manifestly covariant under the PBH transformations, because
it is proportional to the definition of $\psi^{(2)}$ for $\alpha=d/2-2$, \ie
\beqa
\cT&=&-\ \frac{8}{(d-2)} \phi^{(0)}\psi^{(2)}.\nonumber
\eeqa
Finally from the formula \reef{ADALd2}, the total conformal anomaly becomes
\beqa
\cA_{t}&\!\!\!\!=\!\!\!\!&-\ \frac{16}{d} \phi^{(0)}\psi^{(2)},\nonumber
\eeqa
where $\psi^{(2)}$ is given in \reef{Psi2}. This is the anomaly formula for the scalar field which depends on the coordinates of the boundary spacetime.
\subsection{n=2, k=1, general d} \label{n2k1}
In this section, we aim to find the anomaly for four scalars with two derivatives. In the case of $k=1$ and $n=2$, the value of $\al$ is specified by considering \reef{alfa} as $\al=(d-2)/4$. The applicable expansions of the scalar field and of the metric are as follows:
\beqa
\phi^{(\al=\frac{d}{4}-\frac{1}{2})}&=&\phi^{(0)}+\rho \ \phi^{(1)}+\ldots+\rho ^{\al} \Big(\phi^{(\al)}+\rho \phi^{(\al+1)}+\rho \log \rho \psi ^{(\al+1)}+\ldots \Big)+\ldots ,\nonumber\\
{g}^{(\al=\frac{d}{4}-\frac{1}{2} )}&=&{g^{(0)}}_{i j}+\rho \  {g^{(1)}}_{i j}+\ldots+\rho ^{\al} \Big({g^{(\al)}}_{i j}+\rho {g^{(\al+1)}}_{i j}+\ldots \Big)\nonumber\\
&+&\rho^{2 \al} \Big(g^{(2\al)}+\rho g^{(2\al+1)}+\rho  \log \rho {h^{(2\al+1)}}_{i j} +\ldots \Big).
\eeqa
Since the action \reef{toaction} is invariant under $\Phi\rightarrow -\Phi$, the components at orders with odd coefficients of $\frac{\al}{2}$ vanish in above expansions.

Here one needs to explore the different elements to build the anomaly \reef{detex}. From the third equation at orders $\alpha -2$, $\alpha-1$ and  $ 2\alpha-2$, we get, respectively
\beqa
\Tr g^{(\al)}&=&-\ \frac{4 d  \pi G_N^{d+1}}{d-1}(\phi^{(0)})^2,\nonumber\\
\Tr g^{(\al+1)}&=&\frac{\alpha}{ (1+\alpha )}\Tr (g^{(1)}g^{(\al)})-\frac{8 \pi G_N^{d+1}  (-2+d+d \alpha )}{(d-1) (1+\alpha )}\phi^{(0)}\phi^{(1)},\nonumber\\
\Tr g^{(2\al)}&=&\frac{ (-2+3 \alpha )}{4  (-1+2 \alpha )} \Tr (g^{(\al)}g^{(\al)})+\frac{4 \pi G_N^{d+1}   (d+2 \alpha -3 d \alpha )}{ (d-1) (-1+2 \alpha )} \phi^{(0)}\phi^{(\al)},\eeqa
and finally at order $2\alpha-1$, we obtain
\beqa
\Tr g^{(2\al+1)}&\!\!\!\!=\!\!\!\!&\frac{1}{2  \alpha(d-1)  (1+2 \alpha )}\bigg((d-1)\Big(- (1+4 \alpha ) \Tr h^{(2\al+1)}+4\alpha ^2 \Tr (g^{(1)}g^{(2\al)})\nonumber\\
&&\left.+\alpha  \big(-3 \alpha
\Tr (g^{(1)}g^{(\al)}g^{(\al)})+ (1+3 \alpha )\Tr ( g^{(\al)}g^{(\al+1)}) \big)\Big)-8 \pi G_N^{d+1} \right.\nonumber\\
&&\left.\alpha  \Big(\big(-2 (3+\alpha )+d (5+3 \alpha )\big) \phi^{(1)} \phi^{(\al)}+ (-2+d-2 \alpha +3 d \alpha ) \phi^{(0)}\phi^{(\al+1)}\right.\nonumber\\
&&\left.+2 (d-1) \phi^{(0)}\psi^{(\al+1)}\Big)\bigg),\right.
\eeqa
where we used the component of the inverse metric given in \reef{GALPHA2} and also
\beqa
{G_{(\alpha+1)}^{(-1) i j}}&=&- {g^{(0)i k} } g^{(\al+1)}_{k l}{g^{(0)l j}}+{g^{(0)i m} } g^{(1)}_{m n}  {g^{(0)n k} }  g^{(\al)}_{k l}{g^{(0)l j} } +{g^{(0) i k}} g^{(\al)}_{k l}{g^{(0)l m} } g^{(1)}_{m n}  {g^{(0)n j}}. \nonumber
\eeqa
Furthermore at this order, the coefficient of $\log \rho$ gives us
\beqa
\Tr h^{(2\al+1)}&\!\!\!\!=\!\!\!\!&  -\ \frac{4 \pi G_N^{d+1} \big(d+3 d \alpha -2 (1+\alpha )\big) }{(d-1) (1+2 \alpha )}\phi^{(0)}\psi^{(\al+1)}.\labell{TRH2AL12}
\eeqa
From the scalar equation at orders $\alpha-1$ and $\alpha$, one gets
\beqa
\phi^{(\al)} &=& \frac{\alpha}{2 (d-4 \alpha )}\Tr g^{(\al)}\phi^{(0)},\labell{PHIAL2}\\
\psi^{(\al+1)} &=& -\ \frac{1}{16 (2+d)} \bigg(16\ \Box \phi^{(\al)}+ (d^2-4 )\Big( \Tr g^{(\al+1)} \phi^{(0)}- \Tr (g^{(1)}g^{(\al)})\phi^{(0)}\Big)\nonumber\\
&+&(d-2) (d+6) \Tr g^{(\al)} \phi^{(1)} +12 d \ \Tr g^{(1)} \phi^{(\al)} -8  \big(3  \Tr g^{(1)} \phi^{(\al)} -\partial^i\phi^{(0)}
\partial_i\Tr g^{(\al)} \nonumber\\&+&2 \partial_i g_{(\al)}^{i j}\partial_j\phi^{(0)}+2 \partial_i \partial_j\phi^{(0)}g_{(\al)}^{i j} \big) \bigg),\nonumber
\eeqa
and after simplification, we find
\beqa
\psi^{(\al+1)} &\!\!\!\!=\!\!\!\!&\frac{1}{16 (d-1)^2 (d+2)}\bigg(-16 (d-1)^2 \partial^k \partial_k\phi^{(\al)}+\pi G_N^{d+1}(d-2)\Big(4 (d-1) (16+3 d) \nonumber\\
&&\left.\Box\phi^{(0)} (\phi^{(0)})^2 -(d-2) (8+3 d)  R (\phi^{(0)})^3+32 (d-1) \partial^k(\phi^{(0)})^2\partial_k\phi^{(0)}
\Big)\bigg),\right.
\eeqa
where the first term of the above expression by using the the first formula in \reef{PHIAL2} is a total derivative, so it is removed by adding the appropriate contact term to the action. By using by part integration, we get
\beqa
\psi^{(\al+1)} &\!\!\!\!=\!\!\!\!&\frac{(d-2) (8+3 d) \pi G_N^{d+1}}{16 (d-1)^2 (d+2)}  (\phi^{(0)})^2 \Big(4 (d-1) \Box\phi^{(0)}-(d-2) R\ \phi^{(0)}\Big).\labell{psi445}
\eeqa
It is easy to see that the ratio of the coefficients is the same as what we expect from  \reef{anomalyratio}.
As a next step, from the first equation at orders $\al-1$ and $\al$, one has
\beqa
g^{(\al)}_{i j}&=&-\frac{1}{d-2 \alpha }g^{(0)}_{i j} \Tr g^{(\al)}+\frac{8 \pi G_N^{d+1}  (-d+\alpha ) }{(d-1) (d-2 \alpha )}g^{(0)}_{i j}(\phi^{(0)})^2,\labell{GALGAL13}\\
g^{(\al+1)}_{i j}&=&\frac{1}{(d-1) (1+\alpha ) (2-d+2 \alpha )}\bigg((d-1) \Big(R^{(\al)}_{i j}+2 \alpha  {g^{(1)}}^{k}_{j} g^{(\al)}_{i k}+2 \alpha  {g^{(1)}}^{ k}_{i} g^{(\al)}_{j k}\nonumber\\
 &+&(1- \alpha ) g^{(\al)}_{i j}\Tr \ g^{(1)}\Big)+ \Big((d-1) (1+\alpha ) \big(\Tr\ g^{(\al+1)} -\Tr(g^{(1)}g^{(\al)})\Big)\Big)g^{(0)}_{i j}\nonumber\\
&+&8 \pi G_N^{d+1}  \Big((d-\alpha ) \alpha \phi^{(0)} (g^{(1)}_{i j} \phi^{(0)}+2 g^{(0)}_{i j} \phi^{(1)})-(d-1)\partial_i\phi^{(0)}\partial_j\phi^{(0)} \Big)\bigg).
\eeqa
In above the formulas for $g^{(1)}$ and $\phi^{(1)}$ are given in \reef{g1,g2} \footnote{With the same reason given in section \ref{EOMdk1}, we suppose $\al>1$.} and \reef{pbhfi1}. The formula for $R^{(\al)}_{i j}$ is the same as \reef{Rical} with the corresponding value of $\al$.

The last required element, coming from the order $2\al-1$, is
\beqa
g^{(2\al)}_{i j}&\!\!\!\!=\!\!\!\!&\frac{1}{2 (d-1) (d-4 \alpha )}\bigg((d-1)\Big(-2  \alpha
g^{(\al)k}_{i}
g^{(\al)}_{j k}
 + (-1+\alpha )
 g^{(\al)}_{i j}
 \Tr g^{(\al)}
+g^{(0)}_{i j}\big(-2 \Tr g^{(2\al)}\nonumber\\
&&\left.+ \Tr(g^{(\al)} g^{(\al)})\big)\Big)+8 \pi G_N^{d+1}  (-d+\alpha )  \Big((\phi^{(0)})^2 g^{(\al)}_{i j}+2 g^{(0)}_{i j} \phi^{(0)} \phi^{(\al)}\Big)\bigg),\right.\labell{G2AL5}
\eeqa
where we have used the fact that the Ricci tensor does not receive any component at this order.

The expressions for two parts of the anomaly are
\beqa
\cA_g^{(\al=\frac{d}{4}-\frac{1}{2})}&\!\!\!\!=\!\!\!\!&
\frac{-2d}{16 \pi G_N^{d+1}} \bigg(\frac{1}{2} \Tr g^{(2\al+1)}-\frac{1}{4}\Big(2\Tr (g^{(1)}g^{(2\al)})
+2\Tr (g^{(\al)}g^{(\al+1)})
\Big)+\frac{1}{8}\Big(2 \Tr g^{(1)}\nonumber\\
&&\left.
\Tr g^{(2\al)} +2 \Tr g^{(\al)}\Tr g^{(\al+1)}\Big)-\frac{1}{8}\Big(2\Tr g^{(\al)} \Tr (g^{(1)}g^{(\al)})+\Tr g^{(1)}
\Tr (g^{(\al)}g^{(\al)})
\Big)\right.\nonumber\\
&&\left.+\frac{1}{6}\Big(3 \Tr (g^{(1)}g^{(\al)}g^{(\al)})
\Big)+\frac{1}{48}\Big(3\Tr g^{(1)}\big(\Tr g^{(\al)}\big)^2\Big)\bigg),\right.\nonumber\\
\eeqa
\beqa
\cA_M^{(\al=\frac{d}{4}-\frac{1}{2})}&\!\!\!\!=\!\!\!\!&\frac{-2\alpha (d-\alpha )}{2(d-1)}\bigg(2 \phi^{(1)}\phi^{(\al)} +2 \phi^{(0)}\phi^{(\al+1)} +\frac{1}{2}\Big(2 \Tr g^{(1)}\phi^{(0)}\phi^{(\al)} +2 \phi^{(0)}\phi^{(1)}\nonumber\\
&&\left.
\Tr g^{(\al)} +\Tr g^{(\al+1)} (\phi^{(0)})^2 \Big)-\frac{1}{4}\Big(2\Tr (g^{(1)}g^{(\al)})(\phi^{(0)})^2\Big) +\frac{1}{8}\Big(2(\phi^{(0)})^2\right.\nonumber\\
&&\left.\Tr g^{(1)}\Tr g^{(\al)} \Big)\bigg).\right.
\eeqa
After inserting all the corresponding metric and scalar components, the value of the total anomaly with one derivative and four scalar fields becomes
\beqa
\cA_t^{(\al=\frac{d}{4}-\frac{1}{2})}&\!\!\!\!=\!\!\!\!&- \frac{-2d }{4 (1+2 \alpha )}\Big(\phi^{(0)}\psi^{(\al+1)}
+\frac{(1+4 \alpha ) }{16 \pi G_N^{d+1} \alpha } \Tr h^{(2\al+1)}\Big)=\frac{-2(d+2)}{4 d}\phi^{(0)}\psi^{(\al+1)}.\nonumber
\eeqa
Note that in the first equality, the contributions only come from $\Tr h^{(2\al+1)}$ and $\psi^{(\al+1)}$ and other terms have canceled each other. In the last equality, we have used the relation \reef{TRH2AL12}. The formula for $\psi^{(\al+1)}$ is given in \reef{psi445}.
\subsection{n=2, k=0, general d} \label{n2k0}
Here we wish to determine the anomaly in the case of four powers of the scalar field. Therefore, from the formula \reef{alfa}, one has
$\al=d/4$. The related expansions are
\beqa
{g}^{(\al=\frac{d}{4})}&=&{g^{(0)}}+\rho \  {g^{(1)}}+\ldots+\rho ^{\alpha} \Big({g^{(\al)}}+ \rho {g^{(\al+1)}}+\ldots \Big) +\rho ^{2\alpha}\Big({g^{(2\al)}}+  \log \rho {h^{(2\al)}}+ \ldots\Big)+ \ldots\nonumber\\
\phi^{(\al=\frac{d}{4})}&=& \phi^{(0)}+\rho \ \phi^{(1)}+\ldots+\rho ^{\alpha}\Big(\phi^{(\al)}+ \log \rho  \psi ^{(\al)}+\ldots\Big)+\ldots
\eeqa
Considering the third equation, the formula for $g^{(\al)}$ remains the same as \reef{GALGAL13}. At order $\al-2$, it is manifested that
\beqa
\Tr g^{(2\al)} &\!\!\!\!=\!\!\!\!&\frac{1}{4 (d-1) \alpha  (-1+2 \alpha )}\bigg((d-1) \Big(\alpha  (-2+3 \alpha ) \Tr (g^{(\al)}g^{(\al)})
-2(-1+4 \alpha ) \Tr h^{(2\al)}
\Big)\nonumber\\
&&\left.+16 \pi G_N^{d+1} \alpha \Big((d+2 \alpha -3 d \alpha ) \phi^{(0)}\phi^{(\al)}
-2 (d-1) \phi^{(0)}\psi^{(\al)}
\Big)\bigg),\right.
\eeqa
and the coefficient of the logarithmic term induces
\beqa
\Tr h^{(\frac{d}{2})} &\!\!\!\!=\!\!\!\!& \frac{4 \pi G_N^{d+1}  (d+2 \alpha -3 d \alpha ) }{(d-1) (-1+2 \alpha )}\phi^{(0)}\psi^{(\al)},\labell{TRHD24}
\eeqa
where $\psi^{(\al)}$ is obtained from the scalar equation at order $\al-1$
\beqa
\psi^{(\al)}&=& -\ \frac{d}{16}\Tr g^{(\al)} \phi^{(0)}.\labell{PSIAL4}
\eeqa
The results for the contributions to the anomaly are
\beqa
\cA_ g^{(\al=\frac{d}{4})}&=&\frac{-2d}{16\pi G_N^{d+1}}\Big(\frac{1}{2}\Tr g^{(2\al)} -\frac{1}{4}\Tr (g^{(\al)}g^{(\al)}) +\frac{1}{8}\big(\Tr g^{(\al)}\big)^2 \Big),\nonumber\\
\cA_M^{(\al=\frac{d}{4})}&=&\frac{-2\alpha (d-\alpha )}{2(d-1)}\Big(2\phi^{(0)} \phi^{(\al)}+\frac{1}{2} \Tr g^{(\al)}(\phi^{(0)})^2\Big),
\eeqa
and the total anomaly has the following structure:
\beqa
\cA_t^{(\al=\frac{d}{4})}&\!\!\!\!=\!\!\!\!&-\frac{-2}{16 (d-2) (d-1) \pi G_N^{d+1} }\Big(2 (d-1)^2 \Tr h^{(2\al)}+8 d(d-1) \pi G_N^{d+1}\phi^{(0)}\psi^{(\al)}\nonumber\\
&&\left.+ (\pi G_N^{d+1})^2 d^3 (\phi^{(0)})^4\Big).\right.
\eeqa
However, by using the formulas given in \reef{TRHD24} and \reef{PSIAL4}, in this case, the total anomaly becomes zero. That is, there is not a possibility of receiving a contribution of the four powers of the scalar field in the anomaly.
\subsection{n=3, k=0, general d} \label{n3k0}
We reiterate the discussion in the previous section, in the case of six powers of the scalar field to realize whether there is a contribution to the anomaly for $\al=d/6$. The corresponding expansions are
\beqa
{g }^{(\al=\frac{d}{6})}&=&{g^{(0)}}+\rho \  {g^{(1)}}+\ldots+\rho ^{\al} {g^{(\al)}}+\rho^{2\al} g^{(2\al)}+\rho^{3\al} g^{(3\al)}+\rho ^{3\al}  \log \rho {h^{(3\al)}}+\ldots ,\nonumber\\
\phi^{(\al=\frac{d}{6})}&=& \phi^{(0)}+\rho \ \phi^{(1)}+\ldots+\rho ^{\al} \phi^{(\al)}+\rho ^{2\al} \phi^{(2\al)}+\rho ^{2\al} \log \rho  \psi ^{(2\al)}+\ldots.
\eeqa
We have already obtained some of the above components including $g^{(\al)}$, $g^{(2\al)}$ and $\phi^{(\al)}$ in \reef{GALGAL13}, \reef{G2AL5} and \reef{PHIAL2}, respectively, for the given value of $\al$ in this section.
From the third equation at order $3\al$, one obtains
\beqa
\Tr g^{(3\al)} &\!\!\!\!=\!\!\!\!&\frac{1}{3 (d-1) \alpha  (-1+3 \alpha )}\bigg((d-1) \alpha  (-3+7 \alpha ) \Tr (g^{(2\al)} g^{(\al)} )+\alpha  (-1+d+2 \alpha -2 d \alpha )\nonumber\\
&&\left.
 \Tr (g^{(\al)} g^{(\al)} g^{(\al)} )+(-1+d+6 \alpha -6 d \alpha ) \Tr h^{(3\al)}+4 \pi G_N^{d+1}  \alpha  \Big((d+8 \alpha -9 d \alpha ) \right.\nonumber\\
&&\left.(\phi^{(\al)})^2+2 \big((d+4 \alpha -5 d \alpha ) \phi^{(0)}\phi^{(2\al)}-2 (d-1) \phi^{(0)}\psi^{(2\al)}\big) \Big)\bigg).\right.
\eeqa
Also at the same order, by considering the logarithmic term, we have
\beqa
\Tr h^{(3\al)} &\!\!\!\!=\!\!\!\!& \frac{8 \pi G_N^{d+1}  (d+4 \alpha -5 d \alpha ) }{3 (d-1) (-1+3 \alpha )}\phi^{(0)}\psi^{(2\al)},\labell{TRH3AL6}
\eeqa
where $\psi^{(2\al)}$ according to the order $2\al-1$ of the scalar equation is
\beqa
\psi^{(2\al)} &\!\!\!\!=\!\!\!\!&\frac{d}{48}  \Big(-2 \Tr g^{(2\al)}\phi^{(0)}+\Tr (g^{(\al)} g^{(\al)} )\phi^{(0)} -3 \Tr g^{(\al)} \phi^{(\al)} \Big).\labell{PSI2AL6}
\eeqa
The contributions to the anomaly are coming from
\beqa
\cA_g^{(\al=\frac{d}{6})}&=&
\frac{-2d}{16 \pi G_N^{d+1}}\Big(\frac{1}{2} \Tr g^{(3\al)}-\frac{1}{4}\big( 2\Tr (g^{(\al)} g^{(2\al)}) \big)+\frac{1}{8}\big(2\Tr g^{(\al)}\Tr g^{(2\al)}\big)\nonumber\\
&-&\frac{1}{8} \Tr g^{(\al)}
 \Tr (g^{(\al)} g^{(\al)} ) +\frac{1}{6}\Tr (g^{(\al)} g^{(\al)} g^{(\al)} ) +\frac{1}{48} \big(\Tr g^{(\al)}\big)^3\Big),\nonumber\\
\cA_M^{(\al=\frac{d}{6})}&=&\frac{-2\alpha (d-\alpha )}{2(d-1)}\Big(2\phi^{(0)}\phi^{(2\al)}+(\phi^{(\al)})^2+ \frac{1}{2}\big(\Tr g^{(2\al)} (\phi^{(0)})^2+2 \Tr g^{(\al)}\phi^{(0)}\phi^{(\al)}\big)\nonumber\\&-&\frac{1}{4}  \Tr (g^{(\al)} g^{(\al)})(\phi^{(0)})^2+\frac{1}{8}\big(\Tr g^{(\al)}\big)^2 (\phi^{(0)})^2\Big),
\eeqa
and the total anomaly becomes
\beqa
\cA_t^{(\al=\frac{d}{6})}&\!\!\!\!=\!\!\!\!&-\frac{-2d }{3 (d-2)}\bigg(\phi^{(0)}\psi^{(2\al)}+\frac{3 (d-1)} {8  \pi G_N^{d+1} d}\Tr h^{(3\al)}-\frac{ (\pi G_N^{d+1})^2 d^3}{3 (d-1)^2}(\phi^{(0)})^6\bigg).
\eeqa
The same as before, we see that the anomaly after inserting of the formulas given in \reef{TRH3AL6} and \reef{PSI2AL6} vanishes.
\section{Discussion} \label{discuss}
We have used the AdS/CFT correspondence that prescribes a duality between bulk/boundary observables in order to construct iteratively the bulk solutions from CFT data. We did not have any assumption except that the CFT has an AdS dual.
We then proceeded to investigate logarithmically divergent part of the asymptotic solution to find the conformal anomaly.
All infrared divergences of the bulk on-shell action can be obtained in terms of boundary value of fields.
Our discussion for scalar fields can be straightforwardly generalized to other kinds of matter \cite{Nojiri:2000hr,Uhlemann:1011he}.

However, in general, a knowledge of sources in the CFT is not sufficient to construct the complete bulk metric-scalar solution. By additional information from the expectation value of the dual operators, the undetermined non-local coefficients in each case are obtainable. On the other hand we expect by using the information from the trace and divergence of $g_{(d)}$ one can find the explicit form which satisfies these constraints. Therefore, this way one could obtain a quantity that is directly related to expectation value of the corresponding boundary stress-energy tensor in any dimension d. Also the solution to the deep interior region is accessible by extra CFT data (for instance non-local observables such as Wilson loop, Wilson surfaces, etc).

In our discussion we supposed certain covariant terms appeared in the anomaly depending on the relevant deformation exist in the boundary theory. To solve the equations of motion we only discussed a specific action combining of \reef{graction} and \reef{scaction}. For a general gravitational action which involves a function of the curvature and its covariant derivatives with the possible coupling to any number of matter fields and their derivatives such that $d+1$ diffeomorphism leaves it unchanged, a Weyl transformation of the on-shell action gives the anomaly. There are two independent types of trace anomalies that are related to the terms that are present in the effective action. The type A anomaly which has the Euler density structure in the desired dimension and the type B anomalies that are an Weyl invariant expression constructed of the contraction of the Weyl tensors or covariant derivatives of the Weyl tensor \cite{Imbimbo:2000diff}. In our calculation we get the form of the type A anomaly for a specific choice of dimensions related to the pure gravity part. From the PBH transformation point of view the type B anomalies are not universal in the sense they appear with arbitrary coefficients. However, these coefficients are fixed by using the equations of motion pertaining to the chosen form of the action.

Throughout this paper we only take the Einstein gravity action into consideration. If higher derivative corrections are included in the action such as contractions of Weyl Tensors and combinations of curvature tensors like $R^n$ \cite{Deser:1993gc,Boulanger:04,Boulanger:06}, we anticipate that the result for the anomaly will be modified. That is, while the general structure will be the same, the coefficients will have new values depending on the extra gravitational couplings. However, with existence higher derivative corrections in the action, in order to figure out all the coefficients in the expansions of the metric and scalar field, one needs information of the higher point functions of the stress-energy tensor as well.

In addition in our calculations we found that with a free bulk scalar, the anomaly vanishes for some power of the scalar field with $n>2$.
It is expected by introducing a potential with higher order interactions for the bulk scalar in the action, we will get non-zero contributions of this form to the anomaly. Another extension of the present work would be studying renormalization group flows that are produced by adding a potential for the bulk scalar. One can use the present approach to extend the related discussion to any dimension d. However, we are not yet able to relate our conclusions to the properties of the conformal field theories in various dimensions.
\acknowledgments:
I would like to thank Robert Myers for introducing me to this problem and for his continued support throughout its completion. I particularly thank Janet Hung for clarifying many aspects and for useful discussions. I would like to acknowledge the
Perimeter Institute for Theoretical Physics for hospitality during the
course of this work. This work has been
supported by the Iranian Ministry of Science, Research and Technology.
Research at Perimeter Institute is supported by the Government of
Canada through Industry Canada and by the Province of Ontario through the
Ministry of Research \& Innovation.

\end{document}